\documentclass[11pt,a4paper]{article}
\pdfoutput=1
\usepackage{jheppub}

\usepackage{multirow, graphicx,amssymb,url,mathrsfs,amsmath}
\usepackage{wrapfig,boxedminipage,setspace,subfigure,epsfig}
\usepackage{amsxtra,amstext,latexsym,dsfont,amsfonts}
\usepackage{color}
\usepackage[dvipsnames]{xcolor}
\usepackage{float}
\usepackage{slashed,comment}
\usepackage{contour}

\usepackage{tikz}
\usetikzlibrary{calc,patterns,angles,quotes}
\usetikzlibrary{decorations.pathreplacing,decorations.markings,snakes}

\usepackage{tabularx, array}
\newcolumntype{L}[1]{>{\raggedright\arraybackslash}p{#1}}
\newcolumntype{C}[1]{>{\centering\arraybackslash}p{#1}}
\newcolumntype{R}[1]{>{\raggedleft\arraybackslash}p{#1}}

\DeclareMathOperator{\csch}{csch}
\DeclareMathOperator{\sech}{sech}

\renewcommand{\l}{\lambda}

\newcommand{\dd}{\mathrm{d}}

\usepackage{xparse}
\ExplSyntaxOn
\NewDocumentCommand{\HS}{m}
 {
  \seq_set_split:Nnn \l_tmpa_seq { ~ } { #1 }
  \seq_map_inline:Nn \l_tmpa_seq { \contour{green}{##1} ~ } \unskip
 }
\ExplSyntaxOff

\title{Brickwall One-Loop Determinant: Spectral Statistics \& Krylov Complexity}

\author[a]{Hyun-Sik Jeong,}
\author[b,c]{Arnab Kundu}
\author[a]{and Juan F. Pedraza}

\emailAdd{hyunsik.jeong@csic.es}
\emailAdd{arnab.kundu@saha.ac.in}
\emailAdd{j.pedraza@csic.es}

\preprint{\texttt{IFT-UAM/CSIC-24-174}}

\affiliation[a]{Instituto de F\'isica Te\'orica UAM/CSIC, Calle Nicol\'as Cabrera 13-15, 28049 Madrid, Spain}
\affiliation[b]{Saha Institute of Nuclear Physics, 1/AF Bidhannagar, Kolkata 700064, India.}
\affiliation[c]{Homi Bhabha National Institute, Training School Complex, Mumbai 400094, India.}

\abstract{
We investigate quantum chaotic features of the brickwall model, which is obtained by introducing a stretched horizon ---a Dirichlet wall placed outside the event horizon--- within the BTZ geometry. This simple yet effective model has been shown to capture key properties of quantum black holes and is motivated by the stringy fuzzball proposal. We analyze the dynamics of both scalar and fermionic probe fields, deriving their normal mode spectra with Gaussian-distributed boundary conditions on the stretched horizon. By interpreting these normal modes as energy eigenvalues, we examine spectral statistics, including level spacing distributions, the spectral form factor, and Krylov state complexity as diagnostics for quantum chaos. Our results show that the brickwall model exhibits features consistent with random matrix theory across various ensembles as the standard deviation of the Gaussian distribution is varied. Specifically, we observe Wigner-Dyson distributions, a linear ramp in the spectral form factor, and a characteristic peak in Krylov complexity, all without the need for a classical interior geometry. We also demonstrate that non-vanishing spectral rigidity alone is sufficient to produce a peak in Krylov complexity, without requiring Wigner-Dyson level repulsion. Finally, we identify signatures of integrability at extreme values of the Dirichlet boundary condition parameter.}

\begin{document}
\maketitle

%
\section{Introduction}\label{}

\subsection*{The big picture: a primer on quantum chaology}
Thermal physics in quantum systems is closely linked to the concept of quantum chaos, a multifaceted and non-unique notion encompassing {\it a priori} inequivalent mathematical ideas and definitions. This broad range of concepts is often referred to as `quantum chaology,'\footnote{This terminology is credited to Michael Berry~\cite{MichaelBerry_1989}.} underscoring the need to catalog the possibilities and features associated with various definitions of quantum chaos. For a detailed review of related ideas, see {\it e.g.}~\cite{D_Alessio_2016}.\footnote{See also~\cite{1977RSPSA.356..375B, PhysRevLett.52.1} for more in-depth discussions.}

The importance of understanding chaos at the quantum level can hardly be overstated. On one hand, quantum chaos plays a fundamental role in understanding the thermalization of quantum systems in condensed matter physics. On the other hand, these concepts have emerged as powerful and essential tools for studying quantum field theories and the quantum aspects of black holes. This is particularly true since a detailed description of thermalization in such systems necessarily involves quantum notions of chaos.

Quantum chaos becomes quantifiable through different notions at various time scales during the real-time evolution of quantum systems.\footnote{Note that, compared to classical dynamical systems, this represents a qualitative difference. In classical systems, only two time scales generically exist: the Lyapunov time and the Poincaré recurrence time.} Specifically, for a system with $N$ degrees of freedom, the dynamical evolution can be divided into the following parts: a dissipation time scale of ${\cal O}(1)$, a scrambling time scale of ${\cal O}(\log N)$, a Heisenberg time scale of ${\cal O}(e^{N})$, and a Poincaré recurrence time at ${\cal O}(e^{e^N})$. A system with a large number of degrees of freedom and a discrete spectrum yields a very large but finite Heisenberg time, which is essentially the longest significant time scale for a quantum system. Encoded in the long-time dynamics are the UV physics of the system.\footnote{Usually, long-time behavior is associated with the IR behavior of the system. This is intuitive, as in the long-time limit, initial perturbations dissipate and thermalize. However, dynamics at even longer time scales become sensitive to the gap in the quantum spectrum and the UV details of the system. This sensitivity is traditionally captured by the spectral form factor.}

It is therefore natural to explore `quantum chaology' across the various time scales of a quantum system. However, this is a challenging task. Only a handful of {\it chaotic} systems, such as variants of the Sachdev-Ye-Kitaev (SYK) model and certain Random Matrix Theories (RMT), offer sufficient control to explicitly compute these quantities (see {\it e.g.}~\cite{Santos_2010,DAlessio:2016aa,Kobrin:2021aa,Amore:2024ihm} for recent work on this topic). Any physical system that helps provide quantitative estimates of chaos measures is certainly worth investigating. In this article, we will consider one such system, the details of which we will review shortly.

As mentioned earlier, various measures of quantum chaos span different dynamical time scales. For example, Out-of-Time-Order Correlators (OTOCs) quantify the growth of higher-point functions of local operators in a QFT, up to the scrambling time. These correlators establish a chaos bound for the Lyapunov exponent: $\lambda \le 2 \pi T$,~\cite{Maldacena2016} where $T$ is the temperature. The Spectral Form Factor (SFF), on the other hand, provides quantitative measures of chaos up to the Heisenberg time scale. Meanwhile, complexity represents an independent notion of the spreading of quantum information. Quantum complexity can be defined for either a state~\cite{Balasubramanian:2022tpr} or an operator~\cite{Parker:2018yvk}, referred to as Krylov complexity for a state and operator Krylov complexity, respectively. This measure captures the growth of a given state or local operator under time evolution, as it develops non-vanishing overlaps with other states or operators in the corresponding Hilbert spaces. Interestingly, operator growth also encodes information about the Lyapunov exponent, which is defined in terms of the OTOCs. See {\it e.g.}~\cite{Parker:2018yvk, Avdoshkin:2019trj, Avdoshkin:2022xuw, Dymarsky:2019elm, Dymarsky:2021bjq} for related discussions and~\cite{Chapman:2021jbh} for a recent review.\footnote{Note, however, that as demonstrated in these works, for continuum QFTs, operator complexity is kinematically dictated to grow exponentially. Therefore, it becomes unclear whether this measure can probe dynamical features at all. This limitation can be mitigated to some extent by considering a four-point correlation function for CFTs as the inner product in the operator Hilbert space, which can detect a Hawking-Page transition in large-$c$ CFTs in two dimensions; see \cite{Kundu:2023hbk}.}

It is evident that the macroscopic manifestation of chaotic features, {\it e.g.}, the Lyapunov exponent, can emerge from independent measures of chaos. This aligns with the generic characteristics of quantum chaos. For example, it is known that the Dip-Ramp-Plateau feature of the SFF broadly emerges from level correlations.\footnote{For now, this refers to spectral rigidity or level repulsion between distant energy levels in the spectrum. The repulsion between the nearest energy levels determines the Heisenberg time scale. In the SFF, spectral rigidity contributes to the linear ramp, while level repulsion between neighboring energy levels determines the plateau.} In the same spirit, in this article, we will explore a specific aspect of the real-time growth of state complexity and provide evidence for its origin.

Note that operator complexity exhibits exponentially growing behavior over time for chaotic systems, whereas for integrable systems, it does not. Thus, it was conjectured to serve as a detector distinguishing integrable from chaotic quantum dynamics, as proposed in~\cite{Parker:2018yvk}. This distinction can be traced back to the specific behavior of the Lanczos coefficients associated with the Gram-Schmidt orthogonalization of the Gelfand-Naimark-Segal (GNS) Hilbert space. State complexity, on the other hand, exhibits four qualitatively distinct phases over time: ramp, peak, slope, and plateau. The presence of the peak has been conjectured to indicate quantum chaos, as argued in~\cite{Balasubramanian:2022tpr}.

The origin of the peak, however, is not yet completely understood. Based on several examples and considerations, {\it e.g.}, in~\cite{Erdmenger:2023wjg, Camargo:2023eev,Huh:2023jxt,Caputa:2024vrn,Baggioli:2024wbz, Huh:2024lcm}, it is evident that a non-vanishing level correlation is responsible for the peak's existence. However, it remains unclear how strong this correlation needs to be. In particular, it would be interesting to disentangle the roles of spectral rigidity and nearest-energy-level repulsion in a controlled model. This is one of the primary goals of this article.

\subsection*{The brickwall model for black holes: motivations and expectations}

In \cite{Das:2022evy, Das:2023ulz}, a toy model of a quantum black hole was considered, wherein an {\it ad hoc} Dirichlet boundary condition was imposed by hand on any probe field propagating in a black hole geometry.\footnote{This is based on the brickwall model introduced by 't Hooft in \cite{tHooft:1984kcu}. For a preliminary discussion of brickwalls in the context of AdS/CFT, see \cite{Kay:2011np, Iizuka:2013kma}.} This Dirichlet wall\footnote{For terminological completeness, we note that this boundary condition can be referred to as a brickwall, a Dirichlet wall, a stretched horizon, {\it etc.}, which we will use interchangeably.} is also given an {\it ad hoc} freedom to be placed at any constant radial slice outside the event horizon of the geometry. Among its various intriguing aspects, it was observed in \cite{Das:2022evy, Das:2023ulz, Das:2023xjr, Krishnan:2023jqn, Krishnan:2024kzf, Krishnan:2024sle} that when the Dirichlet wall is placed sufficiently close to the event horizon, the corresponding SFF of a probe {\it scalar} field exhibits a linear ramp with a unit slope. Furthermore, it was demonstrated that, although the spectrum shows level correlations, it does not belong to the Wigner-Dyson universality class and thus lacks the standard RMT level repulsion. Consequently, the model provides a quantum mechanical framework that exhibits quantum chaotic features, such as a linear ramp in the SFF, without adhering to the RMT universality class. 

{
The Dirichlet boundary condition is chosen to probe the influence of near-horizon geometry on spectral diagnostics in a controlled and tractable setting. While other boundary conditions could certainly be explored, the Dirichlet case offers a minimal and effective framework for capturing nontrivial features of quantum black holes.
}

The probe scalar sector studied in the aforementioned works can simply be understood as the one-loop determinant (see {\it e.g.}~\cite{Denef:2009kn}) arising from fluctuations of the gravitational background and is therefore naturally suited to capture features of, at least, semi-classically quantized gravitational fields. In this article, we will complement these studies by exploring the features of state complexity in this model and, in particular, examining their relationship to the model's characteristic spectral statistics. Before summarizing our results, let us offer a few additional comments on the motivation behind the model. 

As extensively reviewed in~\cite{Das:2022evy, Das:2023ulz, Das:2023xjr}, the introduction of the Dirichlet wall is motivated by explicit fuzzball solutions in string theory and supergravity. In these solutions, the smooth event horizon is replaced by stringy degrees of freedom and structure at the horizon scale. While examples of such solutions in supergravity are technically quite involved, the much simpler Dirichlet wall model has been shown to capture and exhibit certain intriguing features characteristic of quantum black holes. We will, therefore, consider this model to gain insights into potential features of state complexity that could characterize quantum black holes.
Further, in addition to the scalar sector, we will explicitly consider fermionic fluctuations to assess potential universal behaviors in the model's spectral statistics, state complexity, and their possible interrelation.\footnote{Note that we do not account for any fluctuations of the Dirichlet boundary conditions in the computation of the one-loop determinant. This is a provisional working prescription, as we lack a precise understanding of the UV origin of the Dirichlet wall. Nonetheless, it is possible to consider an Extremely Compact Object, along with its fluctuations, which could yield similar qualitative physics (see {\it e.g.}~\cite{Mathur:2024mvo}). The Dirichlet boundary condition captures the essential physics, and thus we focus on this model.}

A simple yet noteworthy observation is that, by construction, the Dirichlet wall cuts off the black hole's interior, thereby eliminating any classical interior geometry. This, in turn, removes the need for gravitational wormholes, such as the Einstein-Rosen bridge. While earlier proposals of (circuit) complexity relied crucially on the classical black hole interior (see {\it e.g.}\cite{Brown:2015bva, Stanford:2014jda, Susskind:2018pmk}), complexity is, {\it a priori}, an independent concept. Intriguingly, in our framework, the calculation of Krylov state complexity does not require the assumption of a classical black hole interior.\footnote{It has been suggested that the Krylov complexity of chord states in the double-scaled SYK model may be connected to the length of the dual Lorentzian wormhole in Jackiw-Teitelboim gravity~\cite{Rabinovici:2023yex}.} While it is expected that complexity plays a key role in the emergence of spacetime \cite{Czech:2017ryf, Caputa:2018kdj, Pedraza:2021mkh, Pedraza:2021fgp, Pedraza:2022dqi, Carrasco:2023fcj}, understanding the significance (or lack thereof) of the interior geometry in capturing expected features of a quantum black hole remains an important question. In this article, we add another contribution to this ongoing discussion.

\subsection*{Summary of results and outline}

In this study, we build upon prior work in three key directions: (i) We broaden our exploration of quantum chaos features in the model by analyzing a novel indicator, Krylov state complexity~\cite{Balasubramanian:2022tpr}, which is known to correlate with the SFF at late times. (ii) We consider a broader range of random matrix ensembles (GOE, GUE, and GSE), whereas previous investigations focused primarily on the GUE ensemble; (iii) We extend the analysis of the brickwall model to include both scalar and fermionic probe fields.\footnote{Fermionic probes, such as the spectral function, are key observables in strongly correlated materials, experimentally accessible via ARPES or STM. Their importance in holography, particularly regarding potential non-Fermi liquid signatures, has been highlighted in pioneering studies~\cite{Lee:2008xf,Liu:2009dm,Cubrovic:2009ye,Faulkner:2009wj,Iqbal:2009fd};}

We find that, depending on the boundary conditions imposed on the Dirichlet wall, which are associated with the variance $\sigma_0$ of the Gaussian distribution, our results align closely with predictions from RMT for a $\beta$-ensemble that interpolates between GOE, GUE, and GSE and explicitly depends on $\sigma_0$. Specifically, we observe the expected level spacing distributions, the linear ramp in the SFF, and the characteristic peak in Krylov complexity. Additionally, we show that, for the extremal limits of $\sigma_0$ (either large or small), the brickwall model approaches integrability, exhibiting Poisson statistics for $\sigma_0 \gg 1$ and saddle-dominated scrambling for $\sigma_0 \ll 1$. The latter case still displays some features reminiscent of chaotic behavior due to classical instability: the ramp in the SFF and the peak in Krylov complexity, accompanied by mild oscillations in its time evolution. However, the level spacing distribution significantly deviates from both RMT and Poisson distributions. All these results hold true for both the scalar and fermion sectors, indicating that these are robust features of the model.

The structure of this paper is as follows. In Section \ref{sec2}, we provide an overview of traditional spectral diagnostics, including level spacing distributions, the SFF, and Krylov complexity of states. In Section \ref{sec31}, we introduce the brickwall models with both scalar and fermionic fields and compute the associated normal modes. Section \ref{sec32} presents our analysis of various quantum chaos indicators—level spacing distributions, SFF, and Krylov complexity—based on the normal modes obtained in Section \ref{sec31}, and we compare these results with those from random matrix theory. Finally, in Section \ref{sec4}, we present our final remarks and a brief outlook on potential future directions. We relegate some technical details to Appendices \ref{appenA} and \ref{appenB}.

%
\section{Preliminaries}\label{sec2}
In this section, we establish the necessary preliminaries for our study of quantum chaos, which will be central to the main analysis in subsequent sections. We provide a  brief review of the dynamical properties described by random matrix theory (RMT) in the context of quantum chaos, focusing not only on traditional spectral diagnostics -- such as level spacing distributions and the spectral form factor -- but also on more recent approach, the Krylov complexity of states.

\subsection{Random matrix theories, level spacing, and spectral form factor}\label{}
RMT plays a pivotal role in identifying universal features of quantum chaotic systems, providing a statistical framework for understanding complex quantum dynamics. A central conjecture in the study of quantum chaos posits that the fine structure of the energy spectrum of a quantum chaotic Hamiltonian can be well approximated by the statistical behavior of random matrices~\cite{Dyson:1962es,Bohigas:1983er}. For a more comprehensive review, see~\cite{Guhr:1997ve,MEHTA1960395,Dyson:1962oir}.

In the context of this work, we consider the three fundamental universality classes of RMT -- the Gaussian Unitary Ensemble (GUE), Gaussian Orthogonal Ensemble (GOE), and Gaussian Symplectic Ensemble (GSE) -- each corresponding to distinct symmetry properties of the Hamiltonian. The classification into these ensembles is determined entirely by the symmetries of the system, with each ensemble characterized by a specific distribution of matrix elements governed by a Gaussian measure. This symmetry-based classification underscores the universality of RMT, making it an essential framework for capturing the statistical behavior of quantum chaotic systems.

\paragraph{The level spacing distribution.}
One of the key signatures of quantum chaos in RMT is the phenomenon of level repulsion, which can be quantitatively described by the level spacing distribution. For chaotic systems, the probability $p(s)$ of finding two adjacent energy levels with a normalized spacing $s$ (i.e., with the mean level spacing set to $1$) follows different distributions depending on the ensemble~\cite{Wigner_1951, Dyson_1962}:
\begin{align}\label{RMTLSD1}
\begin{split}
p_{\text{\tiny{GOE}}} = \frac{\pi}{2} s \, e^{-\frac{\pi}{4}s^2} \,,\qquad 
p_{\text{\tiny{GUE}}} = \frac{32}{\pi^2} s^2 \, e^{-\frac{4}{\pi}s^2} \,, \qquad 
p_{\text{\tiny{GSE}}} = \frac{2^{18}}{3^{6} \pi^3} s^4 \, e^{-\frac{64}{9\pi}s^2} \,.
\end{split}
\end{align}
Level repulsion signifies that, in chaotic systems, energy levels tend to avoid clustering. This is in stark contrast to integrable systems, where the energy levels can accumulate without repulsion. In integrable systems, the spacing between levels follows a Poisson distribution, given by:
\begin{align}\label{RMTLSD2}
\begin{split}
p_{\text{\tiny{Poisson}}} = e^{-s} \,,
\end{split}
\end{align}
which reflects the absence of correlations between energy levels. Thus, while chaotic systems exhibit strong correlations leading to level repulsion, integrable systems are marked by uncorrelated levels that can be arbitrarily close to one another. This difference is a fundamental distinguishing feature between quantum chaotic and integrable systems.

\paragraph{The spectral form factor.}
Another crucial diagnostic of quantum chaos is the spectral form factor (SFF), defined as 
\begin{align}\label{}
\begin{split}
\text{SFF} = \frac{|Z(\beta,t)|^2}{|Z(\beta,0)|^2}\,, \qquad Z(\beta,t) = \text{Tr} \left[e^{-(\beta-i t) H}\right]\,, 
\end{split}
\end{align}
where $\beta,t$ and $H$ are inverse temperature, time, and the Hamiltonian for a given quantum mechanical system, respectively. In the main context with the black hole background with a stretched horizon, we will interpret the normal modes $\omega$ of black hole as eigenvalues of a quantum mechanical system and define the SFF using them, i.e.,
\begin{align}\label{SFFFORBW}
\begin{split}
Z(\beta,t) = \sum_{\omega} e^{-(\beta-i t) \omega}\,.
\end{split}
\end{align}
The SFF offers valuable insights into the time-dependent characteristics of the spectrum. A hallmark of chaotic systems is the emergence of a linear ramp in the SFF at late times, which reflects the onset of universal behavior consistent with RMT~\cite{Brezin:1997aa,Cotler:2016fpe}. In this work, we find a ``linear ramp" as having a slope of approximately $\approx 1$ on a log-log plot. It is noteworthy that while a ramp with a constant slope on a log-log plot indicates a structured behavior, any slope differing from unity should be considered non-linear.

\subsection{Krylov complexity for states}\label{}
In addition to traditional spectral diagnostics, we employ Krylov complexity $C(t)$, or spread complexity, as a modern tool for probing quantum chaos~\cite{Balasubramanian:2022tpr}.\footnote{Krylov complexity, originally introduced for operator growth in the Heisenberg picture~\cite{Parker:2018yvk}, has been adapted to the Schr\"{o}dinger picture to quantify the spread of quantum states in Hilbert space~\cite{Balasubramanian:2022tpr}.} Krylov complexity quantifies the spread of a quantum state over the Krylov basis, generated through the Lanczos algorithm using the system's Hamiltonian, offering a complementary perspective to spectral measures and enhancing our understanding of chaotic behavior.

Krylov complexity evolves over time by incorporating information from the Lanczos coefficients ($a_n, b_n$). While the physical interpretation of these coefficients remains somewhat ambiguous, Krylov complexity is conjectured to serve as a robust probe for chaotic systems, particularly in connection with RMT~\cite{Balasubramanian:2022tpr}. Especially, for time-evolved thermofield double (TFD) states, Krylov complexity reveals a four-stage behavior characterized by a linear ramp, a \textit{peak}, a downward slope, and a plateau. The presence of the peak is posited as indicative of chaotic dynamics within the system. 

This four-stage behavior is analogous to the slope, dip, ramp, and plateau observed in the SFF~\cite{Balasubramanian:2022tpr, Erdmenger:2023wjg}. Furthermore, it has been demonstrated that a direct relationship exists between the saturation value of Krylov complexity and the SFF at late times when $\beta=0$~\cite{Cotler:2016fpe,Erdmenger:2023wjg,Rabinovici:2020ryf,Rabinovici:2022beu}:
\begin{align}\label{SFFCRE}
   {\lim_{T\rightarrow \infty } \frac{1}{T}\int_{0}^{T}\text{SFF}(t) = \frac{1}{d} =\frac{1}{1+2C(t=\infty)} \,,}
\end{align}
where $d$ is the system size of Hamiltonian, and the late-time behavior of $C(t)$ is used
\begin{align}\label{LATECOMAPP}
    C(t=\infty) \approx \frac{d - 1}{2}\,,
\end{align}
which holds for a maximally-entangled state such as the TFD state with $\beta = 0$~\cite{Erdmenger:2023wjg}.

Beyond its application to quantum chaos, Krylov complexity has been widely explored in diverse fields, including RMT, topological and quantum phase transitions~\cite{Caputa:2022eye,Afrasiar:2022efk,Caputa:2022yju,Pal:2023yik}, conformal field theories~\cite{Dymarsky:2019elm,Dymarsky:2021bjq,Kundu:2023hbk,Malvimat:2024vhr}, saddle-dominated
scrambling~\cite{Bhattacharjee:2022vlt,Huh:2023jxt}, and open quantum systems~\cite{Bhattacharya:2022gbz,Bhattacharjee:2022lzy,Mohan:2023btr,Bhattacharya:2023zqt,Bhattacharjee:2023uwx,Carolan:2024wov}, and other related contexts~\cite{Barbon:2019wsy,Yates:2021asz,Caputa:2021ori,Patramanis:2021lkx,Trigueros:2021rwj,Rabinovici:2020ryf,Rabinovici:2021qqt,Rabinovici:2022beu,Bhattacharya:2023xjx,Bhattacharjee:2022qjw,Chattopadhyay:2023fob,Bhattacharjee:2023dik,Bhattacharjee:2022ave,Takahashi:2023nkt,Camargo:2022rnt,Avdoshkin:2022xuw,Erdmenger:2023wjg,Hashimoto:2023swv,Camargo:2023eev,Iizuka:2023pov,Caputa:2023vyr,Fan:2023ohh,Vasli:2023syq,Gautam:2023bcm,Iizuka:2023fba,Anegawa:2024wov,Caputa:2024vrn,Chen:2024imd,Caputa:2024xkp,Chattopadhyay:2024pdj,Rabinovici:2023yex,Nandy:2023brt,Aguilar-Gutierrez:2023nyk,Camargo:2024deu,Aguilar-Gutierrez:2024nau,Bhattacharjee:2024yxj,Baggioli:2024wbz,Balasubramanian:2024ghv,Craps:2024suj,Alishahiha:2024vbf,Gill:2024acg,Li:2024ljz,Jha:2024nbl,Huh:2024lcm,Camargo:2024rrj,He:2022ryk,Nandy:2024mml,Li:2024kfm,Zhai:2024odw,Zhai:2024tkz,Xu:2024gfm,Bhattacharya:2024szw}. For a comprehensive review, including detailed references, see~\cite{Nandy:2024htc}.

\paragraph{Krylov complexity.}
We provide a brief review of the Krylov complexity of states as outlined in~\cite{Balasubramanian:2022tpr}. Consider a quantum system characterized by a time-independent Hamiltonian $H$. The unitary time evolution of a state  $\vert \psi(t) \rangle$ is governed by the Schr$\text{\"{o}}$dinger equation:
\begin{align}\label{sch}
    i \partial_t |\psi(t) \rangle = H |\psi(t) \rangle \,.
\end{align}
The solution to this equation can be expressed as $|\psi(t) \rangle=e^{-i H t}|\psi(0) \rangle$, which can be represented as a power series of states as
\begin{align}
    | \psi(t) \rangle = \sum_{n=0}^\infty \frac{(-i t)^n}{n!} | \psi_n \rangle \,, \qquad |\psi_n \rangle := H^n | \psi (0) \rangle \,.
\end{align}
These states span a subspace known as the Krylov space, which is a subspace of the full Hilbert space. However, in general, the states $|\psi_n \rangle$ are not orthogonal or properly normalized. To construct an orthonormal basis within the Krylov space, referred to as the Krylov basis $|K_n\rangle$, we employ the Lanczos algorithm~\cite{Lanczos:1950zz}, which follows the Gram-Schmidt orthogonalization procedure:
\begin{align}\label{Lanczos algorithm}
    | A_{n+1} \rangle = (H - a_n) | K_n \rangle - b_n | K_{n-1} \rangle \,,
\end{align}
where the Lanczos coefficients $a_n$ and $b_n$ are given as
\begin{align}\label{forconclu}
    a_n = \langle K_n | H | K_n \rangle\,, \quad b_n = \langle A_n | A_n \rangle ^{1/2} \,.
\end{align}

The initial conditions and normalization of the Krylov basis are defined by
\begin{align}
    |K_0\rangle = |\psi(0) \rangle \,, \quad | K_n \rangle = b_{n}^{-1} | A_n \rangle \,, \quad b_0 = 0 \,.
\end{align}
Consequently, the Lanczos algorithm can be rewritten as
\begin{align}\label{Lanczos algorithm2}
    H|K_n \rangle = a_n | K_n \rangle + b_{n+1} | K_{n+1} \rangle + b_n | K_{n-1} \rangle \,,
\end{align}
where the Lanczos algorithm terminates when a value of $n$ is reached for which $b_n=0$.
This formulation illustrates that the Hamiltonian acquires a tri-diagonal representation in the Krylov basis
\begin{align}\label{Lanczos algorithm3}
    \langle K_{m}\vert H|K_n \rangle = a_n \delta_{m,n} + b_{n+1} \delta_{m,n+1} + b_n \delta_{m,n-1} \,.
\end{align}

The time-evolved state $|\psi(t) \rangle$ can then be represented using the Krylov basis as
\begin{align}
    |\psi(t) \rangle = \sum_n \psi_n(t) | K_n \rangle\,,
\end{align}
where $\psi_n(t) = \langle K_n | \psi(t) \rangle$ denotes functions satisfying the normalization condition
\begin{align}
    \sum_{n} |\psi_n(t)|^2 = 1\,.
\end{align} 

By applying the results from the Lanczos algorithm \eqref{Lanczos algorithm2}, the Schr\"{o}dinger equation~\eqref{sch} reduces to an effective description of a particle hopping in a one-dimensional chain:
\begin{align}\label{Lanczos sch}
    i\, \partial_t \psi_n(t) = a_n \psi_n(t) + b_{n+1}\psi_{n+1}(t) + b_n \psi_{n-1}(t) \,.
\end{align}
Solving this equation for a given initial state $|\psi(0)\rangle$ with $\psi_n(0) = \delta_{n0}$, Krylov complexity is then defined as
\begin{align}\label{spread complexity}
    C(t)  = \sum_n n |\psi_n(t)|^2 \,,
\end{align}
which quantifies the spreading of the initial state along the one-dimensional chain. For the initial state $\vert \psi(0)\rangle$, we consider TFD state~\cite{Balasubramanian:2022tpr} as 
\begin{align}
    |\psi(0) \rangle = \frac{1}{\sqrt{Z(\beta,t=0)}}\sum_n e^{-\frac{\beta E_n}{2}} | n\rangle\otimes\vert\,n \rangle\,, \qquad H\vert n \rangle=E_{n}\vert n \rangle.
\end{align}

%
\section{Model and results}\label{}

\subsection{Normal mode of BTZ black hole in the brickwall model}\label{sec31}
Let us begin by briefly reviewing the BTZ black hole~\cite{Banados:1992wn,Banados:1992gq}, whose metric, expressed in Schwarzschild coordinates ($t,\,r,\,\varphi$), takes the form
\begin{align}\label{BGBTZ}
\dd s^2 = -f(r) \dd t^2 + \frac{\dd r^2}{f(r)} + r^2 \dd \varphi^2 \,,
\end{align}
where the angular coordinate $\varphi$ is periodic with $2\pi$. The emblackening factor $f(r)$ is determined as a solution of Einstein's equations in three-dimensional gravity with the horizon radius $r=1$:
\begin{align}
f(r) = r^2 - 1 \,.
\end{align}

In the following sections, we will analyze the behavior of fluctuation (probe) fields -- specifically, the scalar field $\Phi$ and the fermionic field $\Psi$ -- in the background geometry given by \eqref{BGBTZ}. Especially, to facilitate the analytical solution of the corresponding equations of motion for these fluctuations, it is convenient to introduce an alternative set of coordinates ($t,\,z,\,\varphi$):
\begin{align}\label{BGBTZ2}
\dd s^2 = -\frac{z}{1-z} \, \dd t^2 + \frac{\dd z^2}{4z(1-z)^2} + \frac{\dd \varphi^2}{1-z} \,,
\end{align}
where the coordinate transformation from \eqref{BGBTZ} to \eqref{BGBTZ2} is
\begin{align}\label{}
r = \sqrt{\frac{1}{1-z}} \,.
\end{align}
In this new coordinates, the AdS boundary is located at $z\rightarrow 1$ and horizon is at $z\rightarrow0$.

It is imperative to note that the equation of motion for the scalar field, Klein-Gordon equation, can also be solved analytically in the original coordinate \eqref{BGBTZ}, for instance see \cite{Das:2022evy, Das:2023ulz, Das:2023xjr}. However, for the fermionic field, this is not the case. The Dirac equation can be solved analytically in the new coordinate \eqref{BGBTZ2}, as demonstrated in \cite{Iqbal:2009fd}.

In this manuscript, we will focus on solving both scalar and fermionic fluctuation equations within the unified framework of the coordinate system \eqref{BGBTZ2}, and  explore the implications for the brickwall model. We will show that the results obtained for the scalar field are consistent with those of \cite{Das:2022evy, Das:2023ulz, Das:2023xjr}, while our results for the fermionic field represent a novel contribution to the literature.

%
\subsubsection{Probe scalar fields}
We now turn our attention to the brickwall model for scalar fields: especially, we follow closely the presentation of \cite{Das:2023ulz}. Our goal is to solve the Klein-Gordon equation analytically
\begin{align}\label{}
(\Box - m_\Phi^2) \Phi = 0 \,,
\end{align}
which can be solved by writing (in the coordinate system \eqref{BGBTZ2})
\begin{align}\label{}
\Phi =  \phi(z) \, e^{-i \omega t } e^{i J \varphi} \,,
\end{align}
with the assuming that $J$ is an integer. The radial part of equation reads
\begin{align}\label{EOMSCA}
\phi''(z) + \frac{\phi'(z)}{z}  + \frac{J^2 z^2 + \omega^2 - z(J^2 + \omega^2 + m_\Phi^2)}{4z^2(1-z)^2} \phi(z)  = 0 \,.
\end{align}
For simplicity, we focus on the massless case, following~\cite{Das:2022evy, Das:2023ulz, Das:2023xjr}. The analytic solution of \eqref{EOMSCA} in terms of hypergeometric functions ${}_2 F_{1}$ is given by 
\begin{align}\label{ANALSOLSCA}
\begin{split}
\phi(z) &= e^{-\frac{\pi \omega}{2}} z^{-\frac{i\omega}{2}} \bigg[ C_1\, e^{\pi \omega}\, _2F_1\left(\frac{i (J-\omega)}{2};\frac{-i (J+\omega)}{2};1-i\omega;z\right)  \\
& \qquad\qquad\qquad + C_2\, z^{i \omega}\, _2F_1\left(\frac{-i (J-\omega)}{2};\frac{i (J+\omega)}{2};1+i\omega;z\right) \bigg]  \,.
\end{split}
\end{align}

Next, to explore the brickwall model, we first analyze this  solution near the AdS boundary ($z\rightarrow1$), allowing us to express $C_2$ in terms of $C_1$:
\begin{align}\label{NORSCALAR}
\begin{split}
\phi_{\text{bdry}}(z) \,\approx\, 
C_1 \, \frac{e^{\pi \omega} \,\Gamma\left[ 1-i\omega \right]}{\Gamma\left[ 1 + \frac{i (J-\omega)}{2} \right]\Gamma\left[ 1 - \frac{i(J+\omega)}{2} \right]} 
+ C_2 \, \frac{\Gamma\left[ 1+i\omega \right]}{\Gamma\left[ 1 - \frac{i (J-\omega)}{2} \right]\Gamma\left[ 1 + \frac{i(J+\omega)}{2} \right]} \,.
\end{split}
\end{align}
Then, we set the normalizability~\cite{Das:2022evy, Das:2023ulz, Das:2023xjr}, $\phi_{\text{bdry}}(1)=0$, leading to 
\begin{align}\label{NSOL}
\begin{split}
C_2 = - C_1 \, e^{\pi\omega} \frac{\Gamma\left[ 1-\frac{i(J-\omega)}{2} \right]\Gamma\left[ 1+\frac{i(J+\omega)}{2} \right]\Gamma\left[ 1-i\omega \right]}{\Gamma\left[ 1+\frac{i(J-\omega)}{2} \right]\Gamma\left[ 1-\frac{i(J+\omega)}{2} \right]\Gamma\left[ 1+i\omega \right]} \,.
\end{split}
\end{align}

In the brickwall model, another boundary condition is imposed at a stretched horizon ($z=z_0$), slightly outside the event horizon ($z=1$). For this purpose, we substitute \eqref{NSOL} into \eqref{ANALSOLSCA} and find the radial solution near the horizon as 
\begin{align}\label{}
\begin{split}
\phi_{\text{hor}}(z) \,\approx\, C_1 \left( P_1 \, z^{-\frac{i\omega}{2}} + Q_1 \,  z^{\frac{i\omega}{2}}  \right) \,,
\end{split}
\end{align}
with
\begin{align}\label{}
\begin{split}
P_1 = 1 \,, \qquad Q_1 =  -\frac{\Gamma\left[ 1-\frac{i(J-\omega)}{2} \right]\Gamma\left[ 1+\frac{i(J+\omega)}{2} \right]\Gamma\left[ 1-i\omega \right]}{\Gamma\left[ 1+\frac{i(J-\omega)}{2} \right]\Gamma\left[ 1-\frac{i(J+\omega)}{2} \right]\Gamma\left[ 1+i\omega \right]}  \,,
\end{split}
\end{align}
where one can check that $|P_1|=|Q_1|$.
It is worth noting that the functional form of $P_1$ and $Q_1$ is simpler in the coordinate system \eqref{BGBTZ2} compared to the original coordinate \eqref{BGBTZ} in \cite{Das:2022evy, Das:2023ulz, Das:2023xjr}.
Then, the boundary condition at the stretched horizon near the event horizon can be given \cite{Das:2023ulz} in the form of
\begin{align}\label{ST0}
\begin{split}
\phi_{\text{hor}}(z=z_0) = C_1 \left( P_1 \, z_0^{-\frac{i\omega}{2}} + Q_1 \,  z_0^{\frac{i\omega}{2}}  \right) =: \phi_0 \,,
\end{split}
\end{align}
which\footnote{Note that the boundary condition \eqref{ST0} is motivated by the angle-dependent profiles that are found in BPS fuzzballs~\cite{Das:2023ulz}.} leads to 
\begin{align}\label{ST1}
\begin{split}
\frac{P_1}{Q_1} = \frac{\phi_0}{C_1 Q_1} z_0^{\frac{i\omega}{2}} - z_0^{i\omega} \,.
\end{split}
\end{align}

To determine the normal modes in the context of the brickwall model, we proceed by introducing phase terms for $P_1$ and $Q_1$:
\begin{align}\label{}
\begin{split}
P_1 = |P_1| e^{i \theta_{\alpha}}  \,, \qquad  Q_1 = |Q_1| e^{i \theta_{\beta}} \,.
\end{split}
\end{align}
Substituting this into \eqref{ST1}, we have
\begin{align}\label{ST2}
\begin{split}
e^{i (\theta_{\alpha}-\theta_{\beta})} = \mu_J \, e^{i \left(\lambda_J \, \omega + \frac{\theta}{2}\right)} - e^{i \theta} \,,
\end{split}
\end{align}
where
\begin{align}\label{}
\begin{split}
\mu_J := \bigg|\frac{\phi_0}{C_1 Q_1}\bigg| \,,  \qquad \lambda_J\,\omega := \text{Arg} \left[\frac{\phi_0}{C_1 Q_1}\right] \,, \qquad \theta = \text{Arg} \left[z_0^{i\omega}\right] \,.
\end{split}
\end{align}
Then, the real and imaginary parts of \eqref{ST2} yield $\mu_J = 2 \cos \left(\lambda_J \, \omega - \frac{\theta}{2}\right)$, as well as the quantization conditions on $\omega$:
\begin{align}\label{ST3}
\begin{split}
\text{Quantization condition:} \quad\, \cos (\theta_{\alpha}-\theta_{\beta}) = \cos (2 \lambda_J\,\omega) \,, \quad \sin (\theta_{\alpha}-\theta_{\beta}) = \sin (2 \lambda_J\,\omega) \,.
\end{split}
\end{align}
These are phase equations that allow us to compute the normal modes $\omega(n\,,J)$, where $n$ is an integer.

Furthermore, we also fix the freedom as $C_1 Q_1 =1$, which leads to $\mu_J = |\phi_0|$ and $\lambda_J\,\omega := \text{Arg} \left[\phi_0\right]$, in other words, the boundary condition at the stretched horizon \eqref{ST0} can be expressed as
\begin{align}\label{}
\begin{split}
\phi_{\text{hor}}(z=z_0) = \mu_J \, e^{i \lambda_J\,\omega} \,,
\end{split}
\end{align}
where $\mu_J$ and $\lambda_J \omega$ can be interpreted as the magnitude and phase of $\phi_0$.\footnote{It has been shown that the redefinition of $\lambda_J$, such as $\lambda_J\rightarrow\lambda_J/\omega$, does not change the physics on the normal modes. See more details in \cite{Das:2023ulz}.}

In essence, the normal mode $\omega(n,J)$ can be determined by solving the quantization conditions \eqref{ST3}, provided we have two crucial pieces of information: (I) the phases ($\theta_{\alpha},\,\theta_{\beta}$), which are derived from the fluctuation equations of motion; (II) the parameter $\lambda_J$. 

Let us discuss further on the second piece, $\lambda_J$.
A key insight from~\cite{Das:2023ulz} is to model  $\lambda_J$ as drawn from a Gaussian distribution.
In particular, by ensuring that $\mu_J=2$ in the zero-variance limit, where $\mu_J = 2 \cos \left(\lambda_J \, \omega - \frac{\theta}{2}\right)$, one can set
\begin{align}\label{}
\begin{split}
\langle \lambda_J \rangle = \frac{1}{2} \log z_0 \,, 
\end{split}
\end{align}
and then vary the Gaussian distribution's variance, denoted by $\sigma_J$, from which $\lambda_J$ is sampled. It is instructive to note that $\langle \lambda_J \rangle$ is heuristically comparable to the position of the stretched horizon. As $z_0 \rightarrow 0$, i.e., as the stretched horizon approaches the event horizon, we find $\langle \lambda_J \rangle\rightarrow-\infty$.

{
Gaussian ensembles provide a natural framework for modeling randomness with analytic control, enabling direct comparison with predictions from random matrix theory. This approach aligns with established methods in the effective description of chaotic systems and random matrix quantum mechanics.
}

In the numerical computation of the normal modes in this manuscript, we primarily set $\langle \lambda_J \rangle=-10^4$, while the variance is taken as $\sigma_J:=\sigma_0/\sqrt{J}$, where $\sigma_0$ is varied as in \cite{Das:2022evy, Das:2023ulz, Das:2023xjr}.
Three important remarks regarding these computations are in order. First, while it is possible to compute the normal modes $\omega(n,J)$ for any integer $n$, we focus on the first non-zero mode $\omega(n=1,J)$, denoted simply as $\omega(J)$ hereafter, in line with previous literature. This is because higher modes are less significant in the statistical analysis of normal mode spectra~\cite{Das:2022evy, Das:2023ulz, Das:2023xjr}. Second, although we have fixed the stretched horizon's location at
$\langle \lambda_J \rangle = -10^4$, we also discuss the dependence of the normal modes on the
location of the brickwall in the Appendix. \ref{appenA}, where we show that features of quantum chaos become more pronounced as the stretched horizon approaches the event horizon. Third, while different choices for the variance, such as $\sigma_J := \sigma_0$, $\sigma_0 / J$, or $\sigma_0 / \sqrt{J}$, are feasible, the essential physics remains unaffected, as discussed in~\cite{Das:2023ulz}.

\begin{figure}[]
 \centering
     \subfigure[$\sigma_0 = 0.025$ (GUE)]
     {\includegraphics[width=6.8cm]{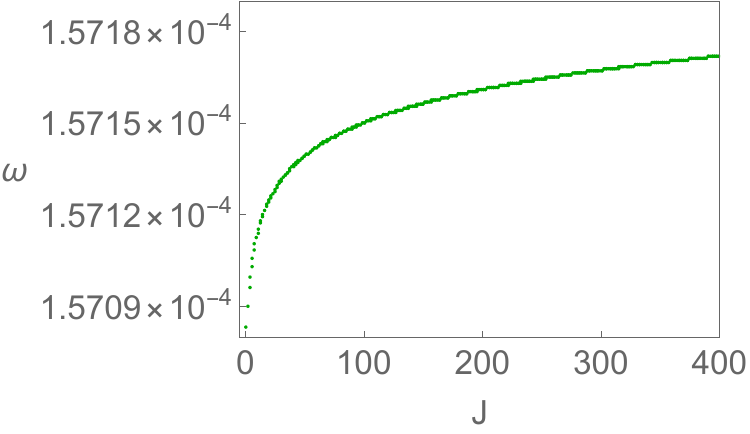} \label{}}
\quad
     \subfigure[$\sigma_0 = 2$ (Poisson)]
     {\includegraphics[width=6.8cm]{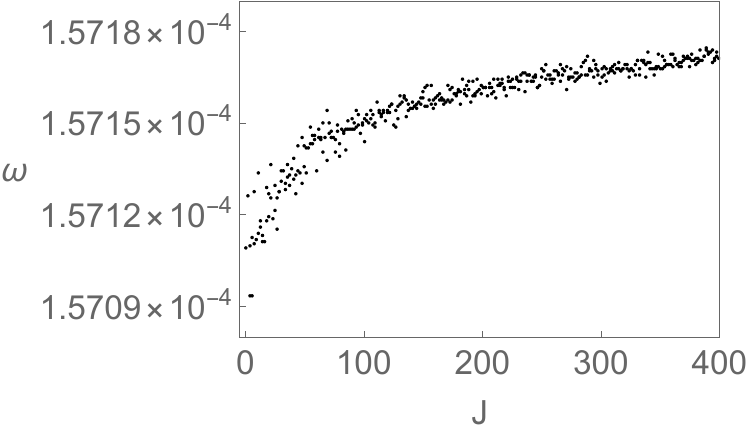} \label{}}
\caption{Normal mode spectrum of scalar fields with $\sigma_0=0.025$ (left), $\sigma_0=2.0$ (right).}\label{NORMALSCALAR}
\end{figure}

In Fig. \ref{NORMALSCALAR}, we present the representative results for the normal mode spectrum of scalar fields, which are consistent with the findings of \cite{Das:2023ulz}.
We observe that as the variance $\sigma_0$ increases, the behavior of the normal modes becomes erratic. In the following section, we will demonstrate that for $\sigma_0 = 0.025$, the level spacing distribution follows GUE statistics, whereas for $\sigma_0 = 2$, it transitions to a Poisson distribution. Additionally, we will examine the dependence of the results on the different value of $\sigma_0$, where the distribution can also be fitted with the GSE and GOE statistics.

%
\subsubsection{Probe fermionic fields}

Next, we discuss the normal mode spectrum of a fermonic field within the brickwall model. Similar to the treatment of the Klein-Gordon equation, our goal here is to solve the Dirac equation analytically, following the formalism outlined in \cite{Iqbal:2009fd}.

We begin with the Dirac equation for a Dirac spinor field $\Psi$ with fermion mass $m_\Psi$:
\begin{equation} \label{eq:DiracEqn}
    \left( \Gamma^M D_M - m_\Psi \right) \Psi = 0 \,,
\end{equation}
where $\Gamma^{M}$ are the gamma matrices and $D_M$ denotes the covariant derivative. It proves useful to employ the coordinate transformation $\rho = \text{arctanh} \sqrt{z}$, where the indices $M, N=(t,\rho,\varphi)$ represent the bulk spacetime and $a, b=(\underline{t},\underline{\rho},\underline{\varphi})$ corresponds to the tangent spacetime, connected via the veinbein $e^a_M$ as 
\begin{equation} \label{MEVER}
    g_{MN} = \eta_{ab} \, e^a_M \, e^b_N \,, \qquad \eta_{ab} = (-1,1,1) \,.
\end{equation}

In terms of the veinbein, the gamma matrices and covariant derivative can be expressed as
\begin{equation} \label{}
    \Gamma^{M} = \Gamma^a e^M_a \,, \qquad D_M = \partial_M + \frac{1}{4} \left( \omega_{ab} \right)_M \Gamma^{ab} \,,
\end{equation}
where the spin connection $\left( \omega_{ab} \right)_M$ and $\Gamma^{ab}$ are defined by  
\begin{equation} \label{MEVERL}
 {
 (\omega^a_b)_M = e^a_N \, e^Q_b \, \Gamma^N_{MQ} - e^Q_b \, \partial_{M} e^a_Q \,,
 \qquad \Gamma^{ab} \, e^M_a \, e^N_b = \frac{1}{2} \left[\Gamma^M, \Gamma^N\right]} \,,
\end{equation}
with the Christoffel symbols $\Gamma^N_{MQ}$.

To solve the Dirac equation \eqref{eq:DiracEqn}, we need to determine two elements: (I) the inverse vielbein $e^M_a$; and (II) the gamma matrices $\Gamma^a$.
Given the metric $g_{MN}$ in \eqref{BGBTZ2}, with $\rho = \text{arctanh} \sqrt{z}$, we choose the diagonal inverse vielbein as
\begin{equation} \label{eq:VielBein}
    e^{t}_{\underline{t}} = \csch{\rho} \,, \qquad e^{\rho}_{\underline{\rho}} = 1 \,, \qquad e^{\varphi}_{\underline{\varphi}} = \sech{\rho} \,,
\end{equation}
along with the gamma matrices:
\begin{equation}\label{eq:Gamma Matrices}
    \Gamma^{\underline{t}}=\begin{pmatrix}
    0 & 1\\
    -1 & 0
    \end{pmatrix}\,,\qquad
    \Gamma^{\underline{\rho}}=\begin{pmatrix}
    1 & 0\\
    0 & -1
    \end{pmatrix}\,,\qquad
    \Gamma^{\underline{\varphi}}=\begin{pmatrix}
    0 & 1\\
    1 & 0
    \end{pmatrix}\,.
\end{equation}

Next, we express the Dirac spinor $\Psi$ in Fourier space\footnote{Note that $\psi_{\pm}$ in \eqref{eq:fermionic field ansatz} are eigenvectors of the gamma matrix $\Gamma^{\underline{\rho}}$ such that $\Gamma^{\underline{\rho}} \, \psi_{\pm}=\pm \psi_{\pm}$.}
\begin{equation} \label{eq:fermionic field ansatz}
\Psi=
    \begin{pmatrix}
    \psi_+(\rho) \\ \psi_-(\rho)
    \end{pmatrix}
    e^{-i \omega t} e^{i J \varphi}\,,
\end{equation}
leading to the following equations of motion for the Dirac spinor:
\begin{equation}\label{eq:DiracEqnfour}
    \left[ \Gamma^{\underline{\rho}} \left( \partial_\rho + \frac{1}{2} \left( \frac{\cosh{\rho}}{\sinh{\rho}} + \frac{\sinh{\rho}}{\cosh{\rho}} \right) \right) + i \left( \frac{J\, \Gamma^{\underline{\varphi}}}{\cosh{\rho}} - \frac{\omega\, \Gamma^{\underline{t}}}{\sinh{\rho}} \right) - m_\Psi \right] \psi = 0\,.
\end{equation}
In order to solve this equation analytically, it is convenient to employ the ansatz within the coordinate system \eqref{BGBTZ2}
\begin{equation}\label{eq:psiTochi}
\psi_\pm(z)=\sqrt{\frac{\left(1\pm\sqrt{z}\right)\sqrt{1-z}}{\sqrt{z}}}\left( \chi_1(z) \pm \chi_2(z) \right)\,, \qquad \rho = \text{arctanh} \sqrt{z} \,,
\end{equation}
where the coordinate $\rho$ is replaced by $z$.

Within the ansatz, the Dirac equations \eqref{eq:DiracEqnfour} reduce to
\begin{align}\label{eq:Dirac Eqn chi12 a}
\begin{split}
    2(1-z)\sqrt{z} \, \partial_z \chi_1 - i\left( \frac{\omega}{\sqrt{z}} + J \sqrt{z} \right) \chi_1 &= \left( m_\Psi-\frac{1}{2}+i(\omega+J) \right)\chi_2\,, \\
    2(1-z)\sqrt{z} \, \partial_z \chi_2 + i\left( \frac{\omega}{\sqrt{z}} + J \sqrt{z} \right) \chi_2 &= \left( m_\Psi-\frac{1}{2}-i(\omega+J) \right)\chi_1 \,,
\end{split}
\end{align}
which admit analytical solutions in terms of hypergeometric functions. Focusing on the massless fermion case, the solutions are
\begin{align}\label{eq:chiSol}
\begin{split}
\chi_1(z) &= (z-1)^{-\frac{1}{4}} \, z^{-\frac{i\omega}{2}} \bigg[ C_1\, z^{i \omega}\, _2F_1\left(\frac{1}{4}-\frac{i (J-\omega)}{2};-\frac{1}{4}+\frac{i (J+\omega)}{2};\frac{1}{2}+i\omega;z\right)  \\
& \qquad\qquad + i \, C_2 \, e^{\pi \omega} \sqrt{z} \,\, _2F_1\left(\frac{1}{4} + \frac{i (J-\omega)}{2};\frac{3}{4}-\frac{i (J+\omega)}{2};\frac{3}{2}-i\omega;z\right) \bigg]  \,, \\
\chi_2(z) &= \frac{2}{1-2i(J+\omega)}\left[ 2(z-1) \, z^{1/2} \, \chi_1'(z) +i (J z +\omega) \, z^{-1/2} \, \chi_1(z) \right] \,.
\end{split}
\end{align}
Then, plugging these expressions \eqref{eq:chiSol} into \eqref{eq:psiTochi} provides the analytic solutions for $\psi_{\pm}(z)$, with two undetermined coefficients, $C_1$ and $C_2$.

Next, expanding these solutions $\psi_{\pm}(z)$ near the AdS boundary ($z\rightarrow1$), we obtain
\begin{align}\label{NBfermion}
\begin{split}
    \psi_+ &\,\approx\, \mathcal{A} (1-z)^{{1}/{2}} \,+\, \mathcal{B} (1-z)  \,+\, \cdots  \,, \\
    \psi_- &\,\approx\, \mathcal{D} (1-z)^{{1}/{2}} \,+\,  \mathcal{C} (1-z) \,+\, \cdots \,.
\end{split}
\end{align}
Here, $\mathcal{A}$ and $\mathcal{D}$ are independent free parameters, while the remaining coefficients, $\mathcal{B}$ and $\mathcal{C}$, are given by the following relations
\begin{equation}\label{CONDBC}
    \mathcal{B} = i(J-\omega) \mathcal{D}\,, \qquad
    \mathcal{C} = - i(J+\omega) \mathcal{A} \,.
\end{equation}
Utilizing the provided analytic solution, the explicit expressions for $\mathcal{A}$ and $\mathcal{D}$ can be obtained as
\begin{align}\label{}
\begin{split}
\mathcal{A} &=   C_1 \frac{(1+i) \,\sqrt{\pi} \,\Gamma\left[ \frac{1}{2} + i \omega \right]}{\Gamma\left[ \frac{1}{4} - \frac{i(J-\omega)}{2} \right] \Gamma\left[ \frac{3}{4} + \frac{i(J+\omega)}{2} \right]} + C_2 \frac{4 (1+i) \sqrt{\pi} \, e^{\pi \omega}\, \Gamma\left[ \frac{3}{2} - i \omega \right]}{(i+2J+2\omega)\Gamma\left[ \frac{1}{4} + \frac{i(J-\omega)}{2} \right] \Gamma\left[ \frac{3}{4} - \frac{i(J+\omega)}{2} \right]}  \,, \\
\mathcal{D} &=   C_1 \frac{(-1)^{1/4} \sqrt{2\pi} \, \Gamma\left[ \frac{1}{2} + i \omega \right]}{\Gamma\left[ \frac{3}{4} - \frac{i(J-\omega)}{2} \right] \Gamma\left[ \frac{1}{4} + \frac{i(J+\omega)}{2} \right]} + C_2 \frac{i (-1)^{1/4} \sqrt{2\pi} \,e^{\pi \omega}\, \Gamma\left[ \frac{3}{2} - i \omega \right]}{\Gamma\left[ \frac{3}{4} + \frac{i(J-\omega)}{2} \right] \Gamma\left[ \frac{5}{4} - \frac{i(J+\omega)}{2} \right]}  \,.
\end{split}
\end{align}

According to the holographic dictionary, it has been demonstrated \cite{Iqbal:2009fd} that the standard fermionic two-point function can be expressed as the ratio of two independent parameters, $\mathcal{D}$ and $\mathcal{A}$, such that $G^{R} = i \mathcal{D}/\mathcal{A}$. This relation implies that normalizability, corresponding to the Dirichlet boundary condition, arises when $\mathcal{A}=0$, analogous to the scalar field case \eqref{NORSCALAR}, where the leading coefficient (i.e., the source term) vanishes. This leads to the following expression: 
\begin{align}\label{C2C1fermion}
\begin{split}
C_2 = - C_1  \frac{e^{-\pi \omega} \left[ \sinh (J \pi) + i \cosh(\pi \omega) \right]}{\pi \,2^{1-2i\omega}} \frac{\Gamma\left[ \frac{3}{2}-i(J+\omega) \right]\Gamma\left[ \frac{1}{2} + i(J-\omega) \right]\Gamma\left[ \frac{1}{2} + i \omega \right]}{\Gamma\left[ \frac{3}{2} - i \omega \right]} \,.
\end{split}
\end{align}

Subsequently, we substitute \eqref{C2C1fermion} into the analytic solution $\psi_{\pm}(z)$ and expaind it near the event horizon ($z\rightarrow0$). We particularly focus on $\psi_{+}(z)$ to analyze the normal mode spectrum in this manuscript, as we impose the Dirichlet boundary condition ($\mathcal{A}=0$), facilitating a comparison with the results for scalar fields under the same boundary condition.\footnote{Notably, in the AdS boundary expansion \eqref{NBfermion}, when $\mathcal{A}=0$, the other function $\psi_{-}$ may effectively resemble a Neumann boundary condition ($\mathcal{C}=0$) via \eqref{CONDBC}, which would produce a different normal mode spectrum.}

Nevertheless, while we checked that the (values of) normal mode spectrum derived from $\psi_{-}(z)$ differs from that of $\psi_{+}(z)$, the spectral statistics (level spacing distribution and spectral form factor) and Krylov complexity remain consistent each other. Thus, it would be practical to focus on $\psi_{+}(z)$, which we will refer to simply as $\psi(z) $ from here on.\footnote{One might inquire whether it is feasible to find common normal modes from both $\psi_{+}$ and $\psi_{-}$. However, our analysis reveals that the quantization conditions preclude the existence of such solutions for $\omega(J)$.}\\

Next, we derive the near event horizon solution as follows
\begin{align}\label{}
\begin{split}
\psi_{\text{hor}}(z) \,\approx\, C_1 \left( P_1 \, z^{-\frac{i\omega}{2}} + Q_1 \,  z^{\frac{i\omega}{2}}  \right) \,,
\end{split}
\end{align}
where
\begin{align}\label{}
\begin{split}
P_1 = -\frac{\cosh(\pi \omega) - i \sinh (J \pi)}{\pi \,2^{-2i\omega}} \frac{\Gamma\left[ \frac{1}{2}-i(J+\omega) \right]\Gamma\left[ \frac{1}{2} + i(J-\omega) \right]\Gamma\left[ \frac{1}{2} + i \omega \right]}{\Gamma\left[ \frac{1}{2} - i \omega \right]} \,, \qquad Q_1 =  1  \,,
\end{split}
\end{align}
and it can be verified that $|P_1|=|Q_1|$.

The subsequent steps to compute the normal mode spectrum follow the same procedure as in the scalar field case:
\begin{align}\label{}
\begin{split}
\psi_{\text{hor}}(z=z_0) = C_1 \left( P_1 \, z_0^{-\frac{i\omega}{2}} + Q_1 \,  z_0^{\frac{i\omega}{2}}  \right) =: \mu_J \, e^{i \lambda_J\,\omega}\,,
\end{split}
\end{align}
with
\begin{align}\label{}
\begin{split}
P_1 = |P_1| e^{i \theta_{\alpha}}  \,, \qquad  Q_1 = |Q_1| e^{i \theta_{\beta}} \,,
\end{split}
\end{align}
where phase parameters $\theta_{\alpha}$ and $\theta_{\beta}$ differ from those in the scalar field case. 

\begin{figure}[]
 \centering
      \subfigure[$\sigma_0 = 0.025$ (GUE)]
     {\includegraphics[width=6.8cm]{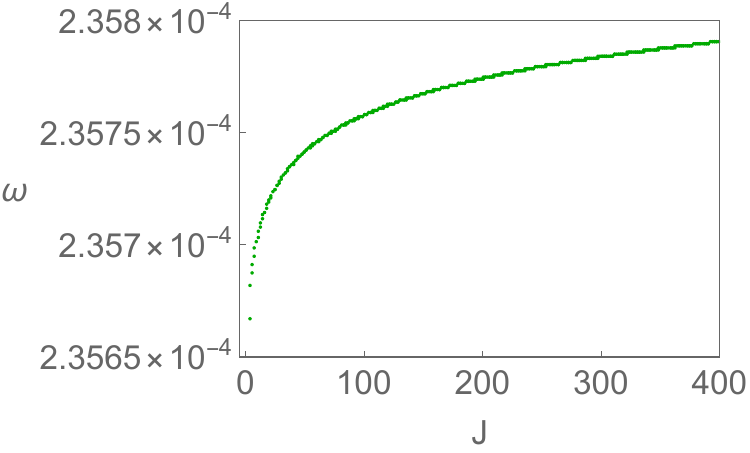} \label{}}
\quad
     \subfigure[$\sigma_0 = 2$ (Poisson)]
     {\includegraphics[width=6.8cm]{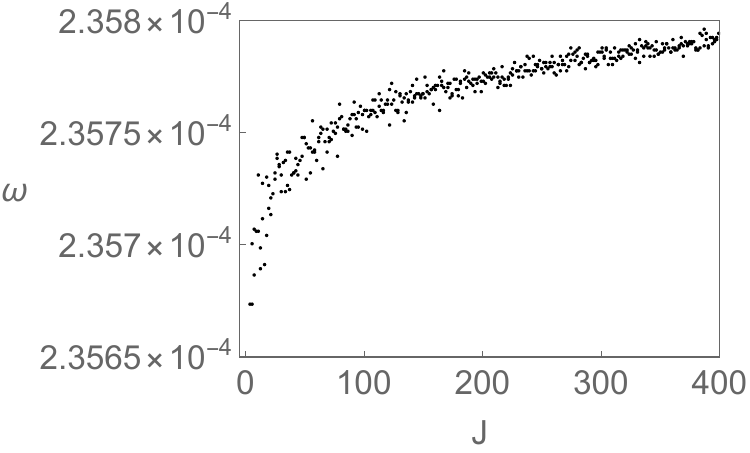} \label{}}
\caption{Normal mode spectrum of fermionic fields with $\sigma_0=0.025$ (left), $\sigma_0=2.0$ (right).}\label{NORMALFERMION}
\end{figure}

By solving the quantization conditions \eqref{ST3} within the same framework, where $\langle \lambda_J \rangle=-10^4$ and $\sigma_J := \sigma_0 / \sqrt{J}$, we obtain the normal mode
spectrum for the fermionic field, as illustrated in Fig. \ref{NORMALFERMION}.
We make two key observations. First, similar to the scalar field case, as the variance $\sigma_0$ increases, the behavior of the normal modes becomes increasingly erratic. Second, the values of the normal modes depend on the nature of the field, whether scalar or fermionic.

In the following section, as in the scalar field case, we will demonstrate that even for the fermionic field, the normal mode spectrum exhibits GUE statistics when $\sigma_0 = 0.025$, while for $\sigma_0 = 2$, it follows a Poisson distribution. One related remark is in order. The observation that the level spacing distribution is \textit{independent} of the type of field may not be surprising. The quantization conditions \eqref{ST3} share a common structure, given by expressions like $\cos(\theta_{\alpha} - \theta_{\beta}) = \cos(2\lambda_J \omega)$ and their sine counterparts. The left-hand side, being a field-dependent function (as both $\theta_{\alpha}$ and $\theta_{\beta}$ vary with the field), provides distinct values for $\omega(J)$, contingent upon whether the field is scalar or fermionic. However, the right-hand side is governed by the same Gaussian distribution through $\lambda_J$. This suggests that while the specific values of $\omega(J)$ differ between the fields, the underlying ``statistical" behavior may remain consistent.

%
\subsection{Dynamical features of random matrices and black holes}\label{sec32}

In this section, we investigate the dynamical properties of random matrices within the framework of brickwall models, utilizing several chaos diagnostics introduced in Sec. \ref{sec2}: specifically, the level spacing distribution, spectral form factor, and Krylov complexity.
Our analysis focuses on the infinite temperature limit, $\beta=0$, as in \cite{Das:2022evy,Das:2023ulz,Das:2023xjr}.\footnote{This limit is particularly advantageous for studying Krylov complexity, as it involves a maximally entangled state with the TFD, thereby clarifying the relationship between the SFF and Krylov complexity \eqref{SFFCRE}.}
All the analysis in the section is grounded in the normal mode spectrum of scalar and fermionic fields derived in the previous Sec. \ref{sec31}, where these modes are interpreted as energy eigenvalues.

\subsubsection{Repulsion in the level spacing distribution}\label{}
\begin{figure}[]
 \centering
     \subfigure[$\sigma_0 = 0.018$ (GSE)]
     {\includegraphics[width=4.83cm]{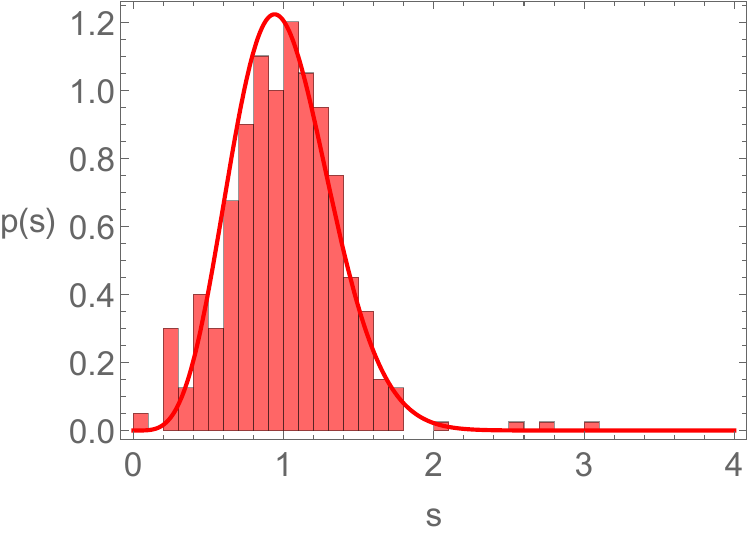} \label{}}
     \subfigure[$\sigma_0 = 0.025$ (GUE)]
     {\includegraphics[width=4.83cm]{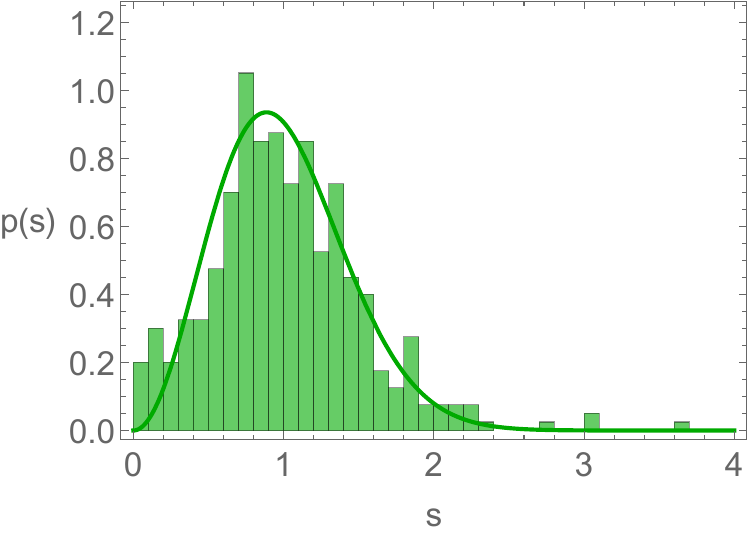} \label{}}
     \subfigure[$\sigma_0 = 0.030$ (GOE)]
     {\includegraphics[width=4.83cm]{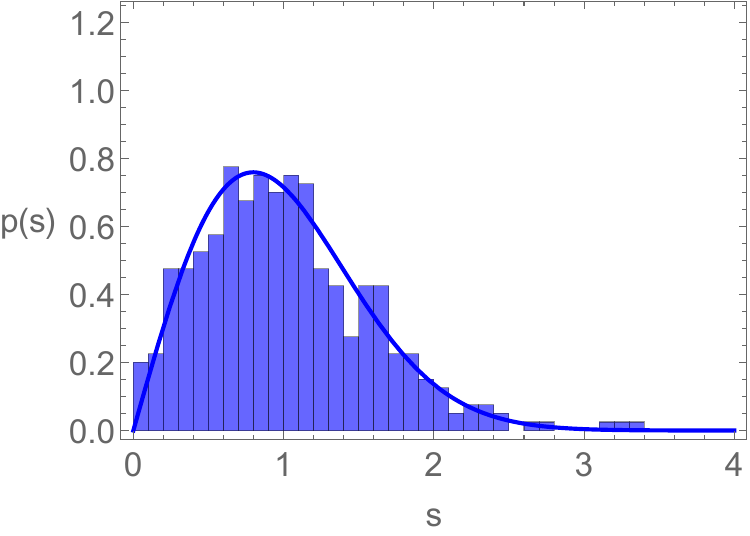} \label{}}
     
     \subfigure[$\sigma_0 = 0.018$ (GSE)]
     {\includegraphics[width=4.83cm]{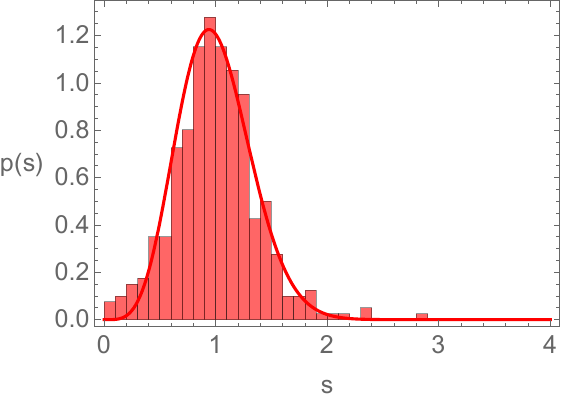} \label{}}
     \subfigure[$\sigma_0 = 0.025$ (GUE)]
     {\includegraphics[width=4.83cm]{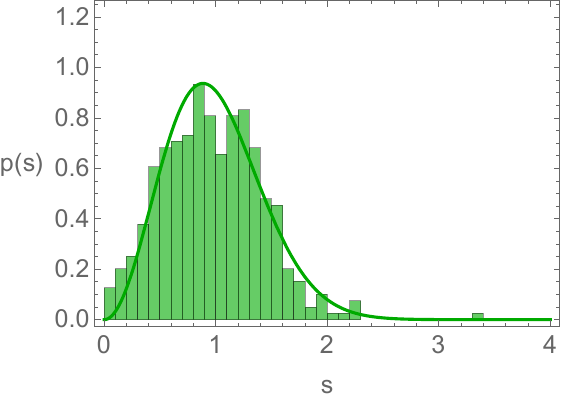} \label{}}
     \subfigure[$\sigma_0 = 0.030$ (GOE)]
     {\includegraphics[width=4.83cm]{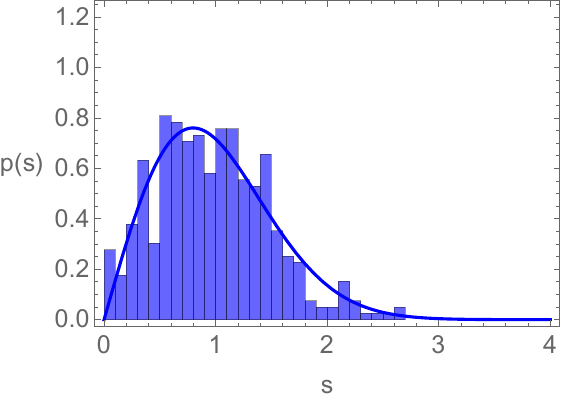} \label{}}     
\caption{Level spacing distributions of scalar fields (top panel), fermionic fields (bottom panel). The solid lines are \eqref{RMTLSD1}.}\label{LSDFIG1}
\end{figure}

In Fig. \ref{LSDFIG1}, we present the level spacing distribution for both scalar fields (top panel) and fermionic fields (bottom panel). 
Our results demonstrate that the brickwall model can produce the level spacing distribution corresponding to all classes of RMT \eqref{RMTLSD1}, when the variance $\sigma_0$ is set to $\sigma_0 = 0.018, 0.025, 0.03$ (GSE, GUE, GOE). Notably, we extend previous analyses of the scalar field case with $\sigma_0 = 0.025$~\cite{Das:2022evy,Das:2023ulz,Das:2023xjr}, which focused exclusively on the GUE case, to encompass all RMT classes for both scalar and fermionic fields. Interestingly, the value of $\sigma_0$ that yields the level spacing distribution appears to be independent of the type of probe field, whether scalar or fermionic. This observation suggests that, despite differences in the normal modes between the fields, the underlying statistical behavior may remain consistent across field types.

\begin{figure}[]
 \centering
     \subfigure[$\sigma_0 = 0$]
     {\includegraphics[width=4.83cm]{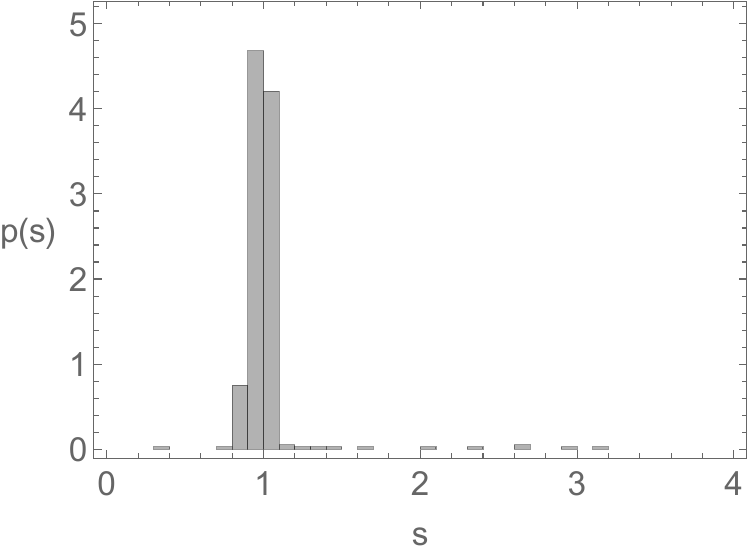} \label{}}
     \subfigure[$\sigma_0 = 2$ (Poisson)]
     {\includegraphics[width=4.83cm]{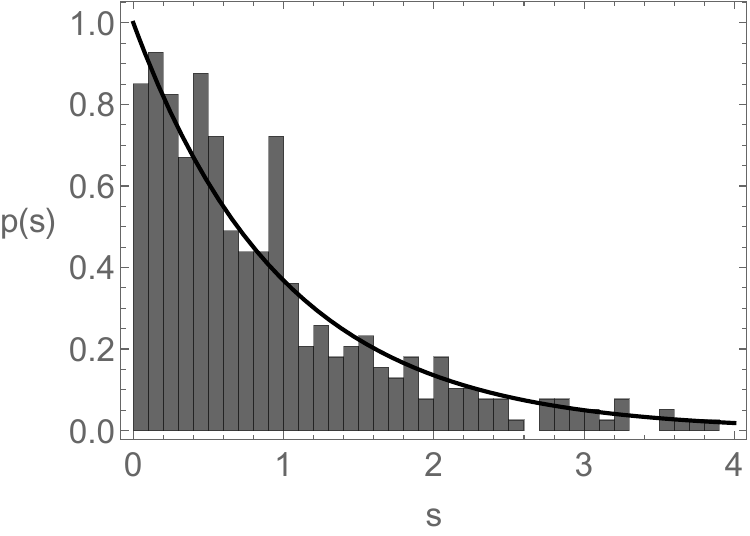} \label{}}
     
      \subfigure[$\sigma_0 = 0$]
     {\includegraphics[width=4.83cm]{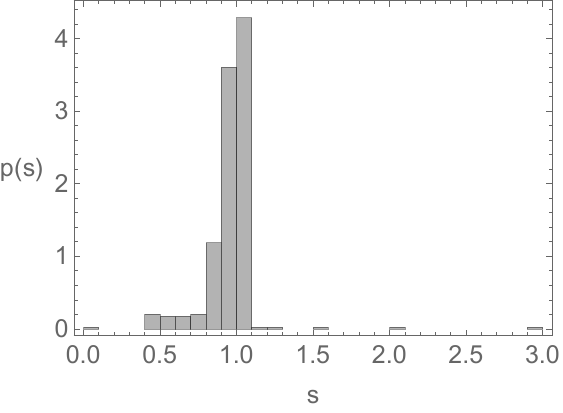} \label{}}
     \subfigure[$\sigma_0 = 2$ (Poisson)]
     {\includegraphics[width=4.83cm]{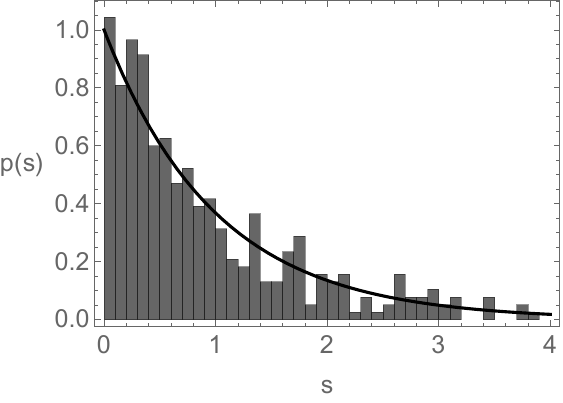} \label{}}
\caption{Level spacing distributions of scalar fields (top panel), fermionic fields (bottom panel). The black solid line is \eqref{RMTLSD2}.}\label{LSDFIG2}
\end{figure}

Additionally, we confirmed the extremal cases reported in \cite{Das:2022evy,Das:2023ulz,Das:2023xjr}, where $\sigma_0$ is either small ($\sigma_0=0$) or large ($\sigma_0=2$). When $\sigma_0=0$, the level spacing distribution collapses to a sharp, almost delta-function-like peak, and as $\sigma_0$ increases to $2$, the distribution transitions to a Poisson distribution \eqref{RMTLSD2} (see Fig. \ref{LSDFIG2}).
In the remainder of this section, we primarily focus on the cases where $\sigma_0 = 0.018, 0.025, 0.03, 2$ (corresponding to GSE, GUE, GOE, and Poisson, respectively), even when examining other probes of chaos. For a more detailed analysis of the dependence of $\sigma_0$ across these regimes, see Appendix \ref{appenB}.\footnote{{In our analysis of level spacing distributions throughout this manuscript, the spectrum has been unfolded in all cases, including the $\sigma_0 = 0$ limit.}}

{
We emphasize that the sensitivity of spectral statistics to small variations in $\sigma_0$ may not be merely a technical detail, but rather a meaningful feature of the model. It suggests that even minimal fluctuations at the stretched horizon can control both the onset of chaotic behavior and the symmetry class of the resulting spectrum. This aligns with well-known transitions observed in quantum chaotic systems and highlights the utility of our setup as a probe of universality and integrability breaking in curved spacetime backgrounds.
}

\subsubsection{Linear ramp in the spectral form factor}\label{}
\begin{figure}[]
 \centering
     \subfigure[$\sigma_0 = 0.018$ (GSE)]
     {\includegraphics[width=4.83cm]{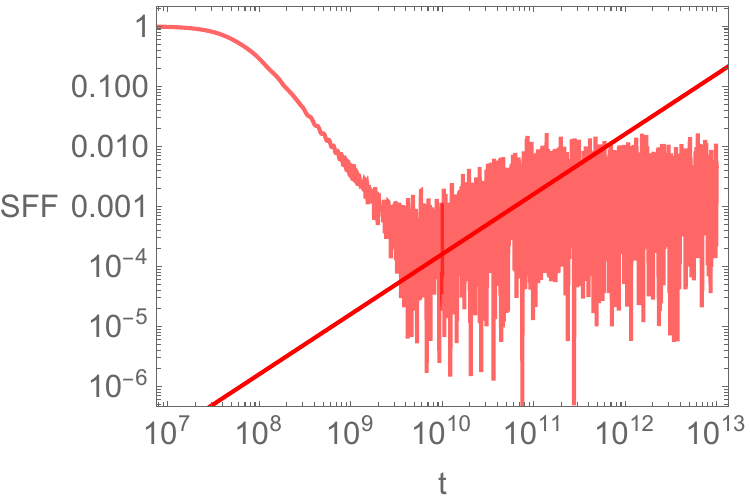} \label{}}
     \subfigure[$\sigma_0 = 0.025$ (GUE)]
     {\includegraphics[width=4.83cm]{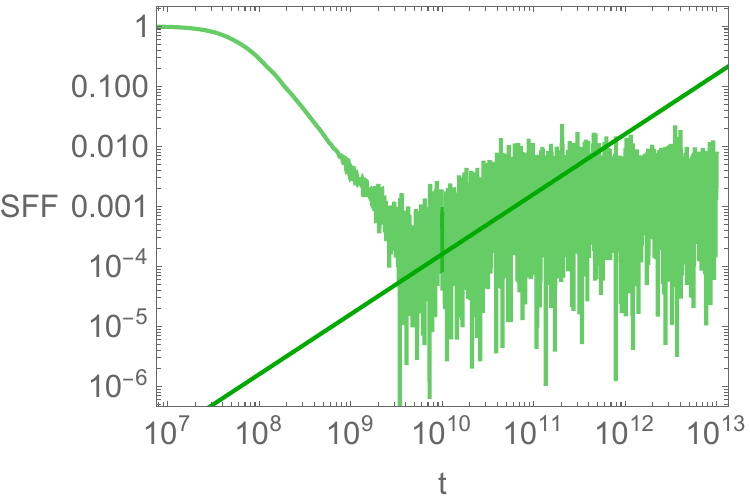} \label{}}
     \subfigure[$\sigma_0 = 0.030$ (GOE)]
     {\includegraphics[width=4.83cm]{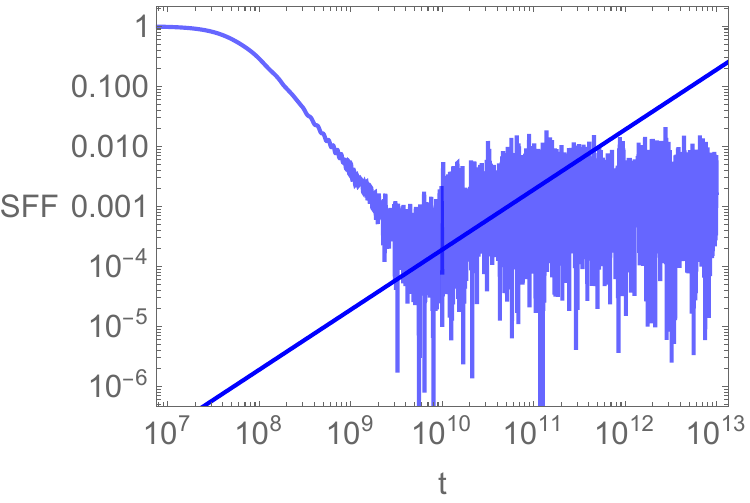} \label{}}
     
     \subfigure[$\sigma_0 = 0.018$ (GSE)]
     {\includegraphics[width=4.83cm]{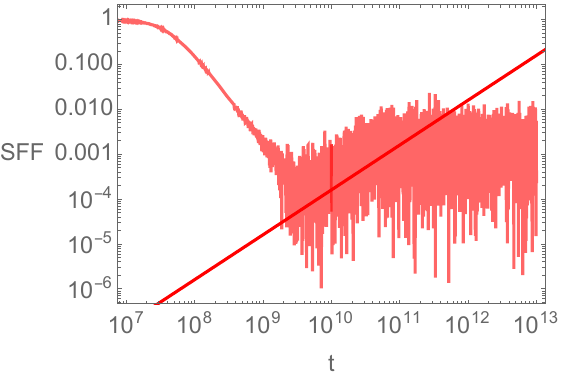} \label{}}
     \subfigure[$\sigma_0 = 0.025$ (GUE)]
     {\includegraphics[width=4.83cm]{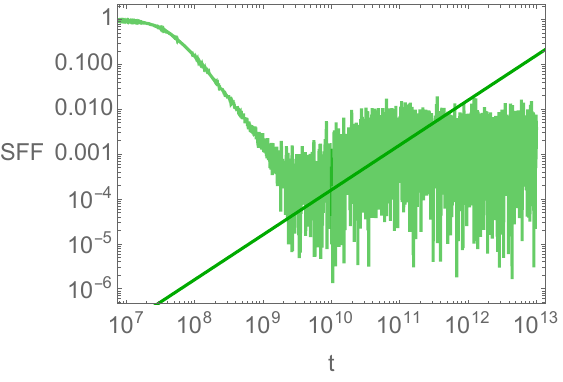} \label{}}
     \subfigure[$\sigma_0 = 0.030$ (GOE)]
     {\includegraphics[width=4.83cm]{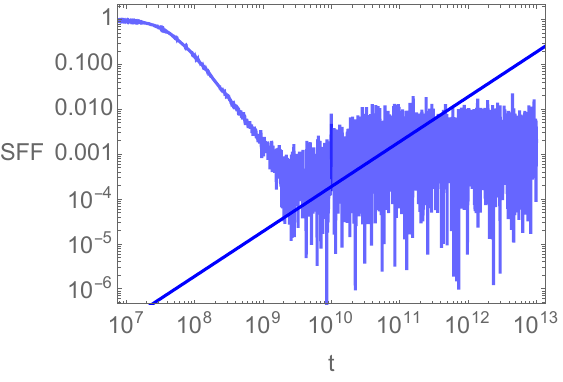} \label{}}
\caption{SFF of scalar fields (top panel), fermionic fields (bottom panel).}\label{SFFFIG1}
\end{figure}

We next turn to the discussion of the spectral form factor (SFF), as defined in \eqref{SFFFORBW}. In Fig. \ref{SFFFIG1}, we present the SFF for both scalar and fermionic fields with $\sigma_0 = 0.018, 0.025, 0.03$, and demonstrate the appearance of a linear ramp, in agreement with the results from RMT~\cite{Brezin:1997aa,Cotler:2016fpe}. The linear slope is also displayed in a log-log plot for further clarity.

\begin{figure}[]
 \centering
     \subfigure[$\sigma_0 = 0$]
     {\includegraphics[width=4.83cm]{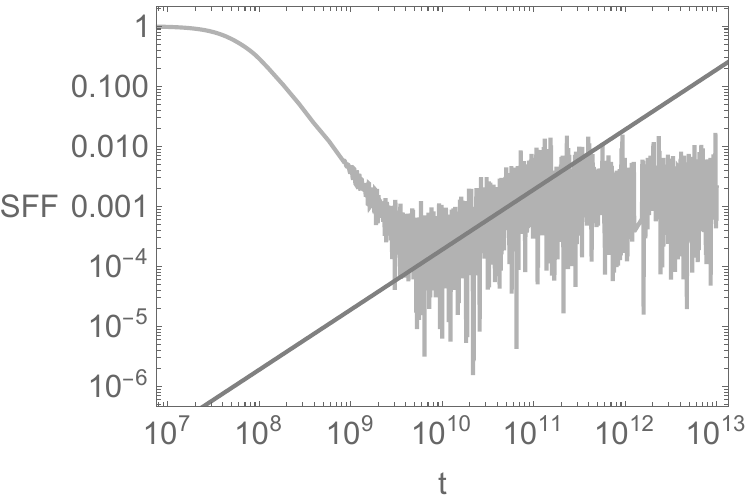} \label{}}
     \subfigure[$\sigma_0 = 2$ (Poisson)]
     {\includegraphics[width=4.83cm]{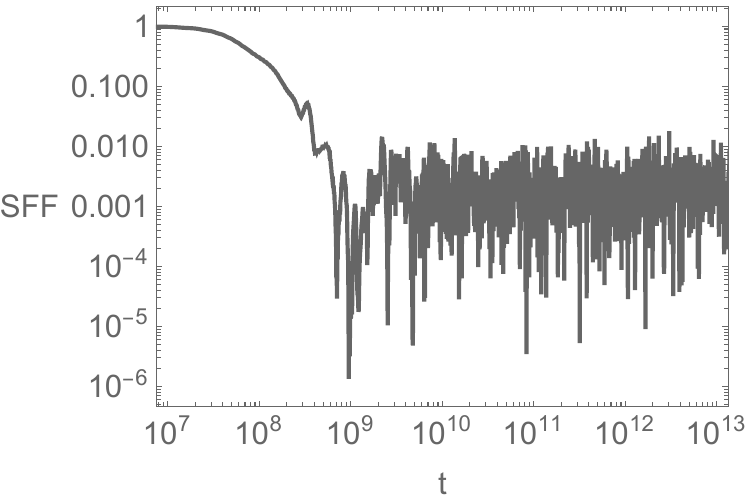} \label{}}
     
     \subfigure[$\sigma_0 = 0$]
     {\includegraphics[width=4.83cm]{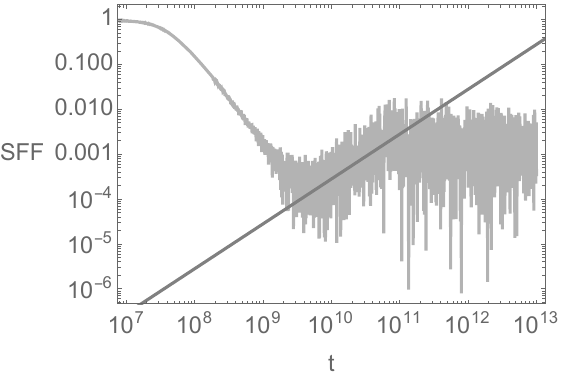} \label{}}
     \subfigure[$\sigma_0 = 2$ (Poisson)]
     {\includegraphics[width=4.83cm]{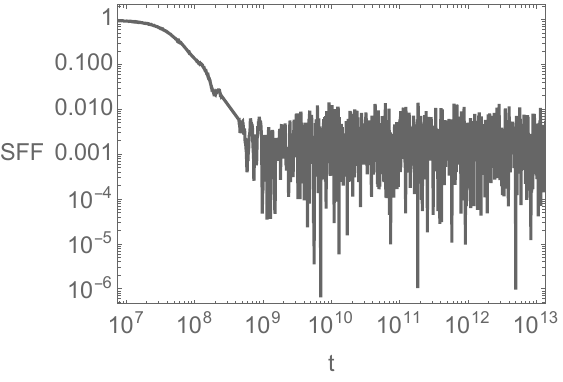} \label{}}
\caption{SFF of scalar fields (top panel), fermionic fields (bottom panel).}\label{SFFFIG2}
\end{figure}

Furthermore, we display the SFF for $\sigma_0 = 0$ and $\sigma_0 = 2$ in Fig. \ref{SFFFIG2}. 
Our findings show that the observations made for the scalar field~\cite{Das:2022evy,Das:2023ulz,Das:2023xjr} also hold for the fermionic field. Specifically, when $\sigma_0 = 0$, the SFF continues to exhibit a linear ramp, while for $\sigma_0 = 2$, where the level spacing follows a Poisson distribution, the SFF no longer displays the ramp. 

It is noteworthy that in the case of $\sigma_0 = 0$, despite the presence of a ramp in the SFF, conventional level repulsion (as seen in GOE, GUE, and GSE) is absent: see Fig. \ref{LSDFIG2} (a) and (c). This suggests that the linear ramp and level repulsion may have independent significance in the context of quantum chaos, rather than being solely connected through random matrix theory. This idea has been further explored in the literature such as through toy models in \cite{Das:2023yfj}, where a simple spectrum $E_n \approx \log n$ can display a linear ramp but do not exhibit traditional level repulsion.

\subsubsection{Characteristic peak of Krylov complexity}\label{}
\begin{figure}[]
 \centering
     \subfigure[Krylov complexity of scalar field]
     {\includegraphics[width=6.5cm]{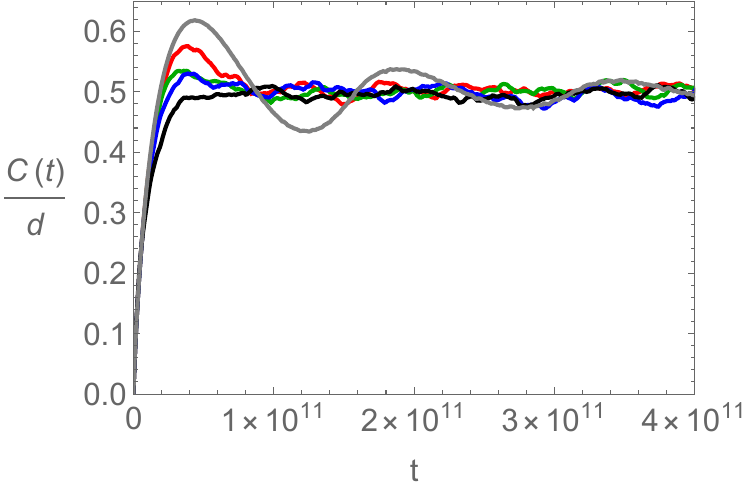} \label{}}
     \subfigure[Krylov complexity of fermionic field]
     {\includegraphics[width=6.5cm]{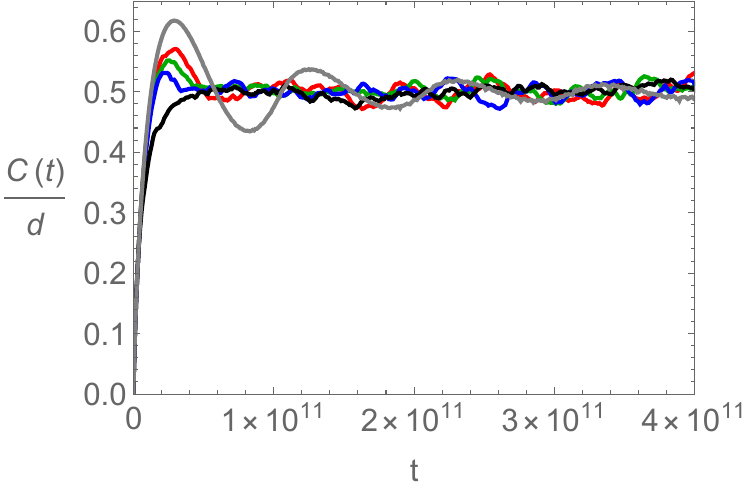} \label{}} 
\caption{Krylov complexity of scalar (left pannel) and fermionic (right pannel) fields when $\sigma_0 = 0,\, 0.018,\, 0.025,\, 0.030,\, 2$ (gray,\, red,\, green,\, blue,\, black). All the Krylov complexity is aligned with the analytic late-time behavior \eqref{LATECOMAPP}.}\label{KryFig1}
\end{figure}

\begin{figure}[]
 \centering
     \subfigure[$a_n$ with $\sigma_0 = 0.018$]
     {\includegraphics[width=4.83cm]{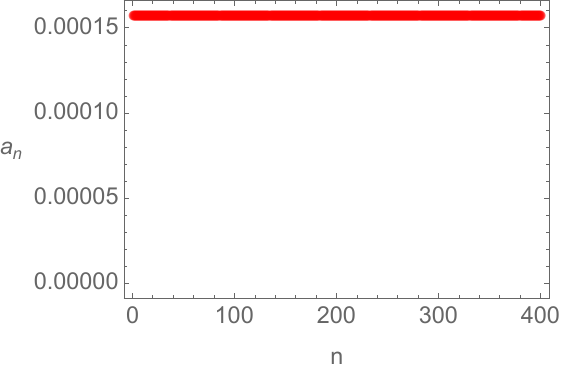} \label{}}
     \subfigure[$a_n$ with $\sigma_0 = 0.025$]
     {\includegraphics[width=4.83cm]{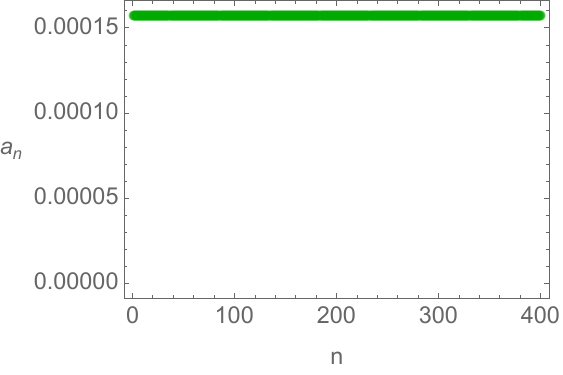} \label{}}
     \subfigure[$a_n$ with $\sigma_0 = 0.030$]
     {\includegraphics[width=4.83cm]{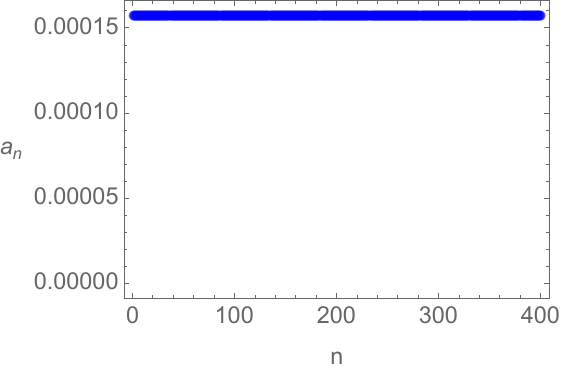} \label{}}
     
     \subfigure[$b_n$ with $\sigma_0 = 0.018$]
     {\includegraphics[width=4.83cm]{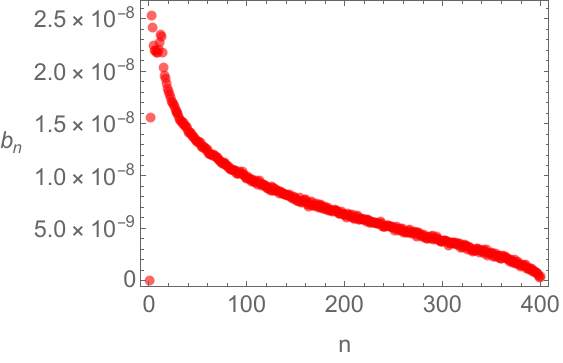} \label{}}
     \subfigure[$b_n$ with $\sigma_0 = 0.025$]
     {\includegraphics[width=4.83cm]{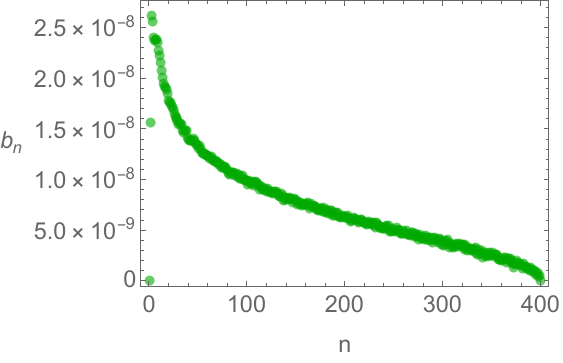} \label{}}
     \subfigure[$b_n$ with $\sigma_0 = 0.030$]
     {\includegraphics[width=4.83cm]{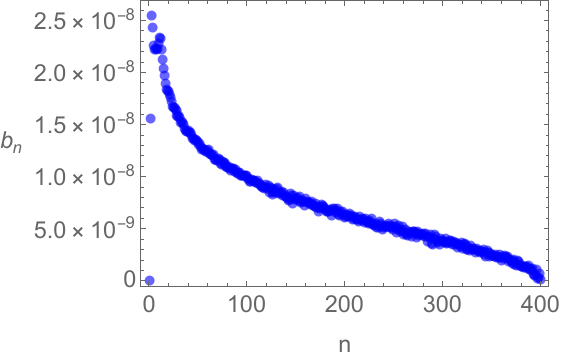} \label{}}
\caption{Lanczos coefficients of scalar fields.}\label{SFFFIGLanc}
\end{figure}

Next, we turn our attention to the Krylov complexity, as defined in \eqref{spread complexity}, derived from the normal modes of scalar and fermionic fields. In Fig. \ref{KryFig1}, we present the Krylov complexity scaled by the system's dimension $d$, specifically $C(t)/d$, which facilitates the examination of its analytical late-time behavior \eqref{LATECOMAPP}: $C(t=\infty)/d \approx 1/2$. 
For our numerical computations, we set $d=400$, as illustrated in Fig. \ref{NORMALSCALAR} and Fig. \ref{NORMALFERMION}.
{We also present the Lanczos coefficients for the scalar field case in Fig. \ref{SFFFIGLanc}.\footnote{Similar behavior is observed for fermionic fields and for other values of $\sigma_0$, showing qualitative agreement with the reported results from random matrix theory and SYK models in the literature.}}

Our findings indicate that for $\sigma_0 = 0.018, 0.025, 0.03$ (GSE, GUE, GOE), the Krylov complexity exhibits a characteristic peak, aligning with previous analyses linking Krylov complexity to RMT~\cite{Balasubramanian:2022tpr}. In contrast, for $\sigma_0=2$ (Poisson), this peak diminishes. Consequently, our analysis within the brickwall model is consistent with the quantum chaos conjecture regarding Krylov complexity. Additionally, we confirm the relationship \eqref{SFFCRE} between the SFF and the Krylov complexity at late times.

Moreover, we observe that when $\sigma_0=0$ (represented by the gray data in Fig. \ref{KryFig1}), the Krylov complexity also displays a peak along with slightly oscillatory behavior. This observation warrants further commentary.

From the analysis conducted for $\sigma_0=0$ in this section, we identify three notable features: (I) the level spacing distribution exhibits delta-function-like behavior, (II) a ramp appears in the SFF, and (III) a characteristic peak manifests in the Krylov complexity. 

All of these features bear resemblance to the results observed in saddle-dominated scrambling scenarios, as studied in well-known models such as the Lipkin-Meshkov-Glick model and inverted harmonic oscillators~\cite{Huh:2023jxt}. Therefore, based on these observations, it is compelling to propose that the brickwall model with $\sigma_0=0$ serves as a promising gravitational toy model for saddle-dominated scrambling, effectively mimicking chaotic features while remaining integrable.

{
Interestingly, we find that a similar pattern emerges in the toy model with logarithmic spectrum $E_n = \log n$, where the level spacing distribution also shows a delta-function-like peak, the SFF exhibits a ramp~\cite{Das:2023yfj}, and the Krylov complexity features a distinct peak with mild oscillations. These shared features further support the structural connection between such models and our brickwall setup with $\sigma_0=0$. While these indicators are not unique to a single system and do not strictly define saddle-dominated scrambling, their concurrent appearance in multiple contexts reinforces the idea that the $\sigma_0=0$ brickwall model may effectively capture some aspects of the regime of saddle-dominated scrambling within a gravitational framework. Nevertheless, a precise justification remains unclear, the interpretation of saddle-dominated scrambling requires further elaboration.
}

%

%
\section{Final remarks and outlook}\label{sec4}

In this work, we have conducted an in-depth investigation into the quantum chaotic features of the brickwall model, constructed by introducing a Dirichlet wall outside the event horizon of a black hole, with a focus on the BTZ geometry. This setup has proven to be an effective framework for probing the quantum properties of black holes, particularly within the contexts of string theory and holography.

Within this geometric setting, we analyzed the dynamics of both scalar and fermionic probe fields. The normal mode spectra of these fields were derived by solving the corresponding Klein-Gordon and Dirac equations, subject to Dirichlet boundary conditions at the wall and normalizability conditions at the asymptotic boundary of AdS.
The resulting normal modes, interpreted as the energy eigenvalues of a quantum mechanical system, were utilized for a detailed analysis of spectral statistics, including the level spacing distribution and the SFF. Additionally, we employed Krylov state complexity~\cite{Balasubramanian:2022tpr}, a novel diagnostic tool for quantum chaos, to gain deeper insights into the chaotic dynamics of the model.\footnote{For a related analysis of Krylov complexity for operators in Schrödinger field theory within the grand canonical ensemble, covering both bosonic and fermionic cases without using the brickwall model, see \cite{He:2024hkw}.}

Our findings demonstrated that, depending on the boundary conditions imposed at the Dirichlet wall ---characterized by the variance $\sigma_0$ of a Gaussian distribution--- the model exhibits behavior consistent with predictions from RMT. Specifically, we observed level spacing distributions interpolating between GOE, GUE, and GSE (depending on the value of $\sigma_0$), alongside a linear ramp in the SFF and a characteristic peak in Krylov complexity, all without the need for a classical interior geometry. Furthermore, at extreme values of $\sigma_0$, the brickwall model also displayed features characteristic of integrable systems, such as Poisson statistics for $\sigma_0 \gg 1$ and saddle-dominated scrambling for $\sigma_0 \ll 1$. {Nevertheless, a precise justification for interpreting this behavior as saddle-dominated scrambling remains unclear and warrants further investigation; see the discussion on page 21 for additional context.}

\vspace{0.2cm}
There are several remarks and directions worth investigating in the future:
\begin{itemize}
\item{At a technical level, our calculations closely align with 't Hooft's brickwall model \cite{tHooft:1984kcu}, which generates a large number of localized quantum states near the horizon, resulting in an entropy proportional to the horizon area. Our findings suggest that such a brickwall scenario may also capture aspects of the fine-grained information of a quantum black hole, raising the important question of which fine-grained quantities can be accessed through semi-classical gravity. Further analysis of the brickwall model would be valuable for exploring these details and understanding how semi-classical methods might serve as a bridge to quantum gravity.}
\item{It is imperative to better understand the boundary conditions on the Dirichlet wall from the perspective of the boundary CFT. The interpretation of the bulk cut-off surface within the dual CFT framework remains unclear, as our construction appears agnostic to any specific UV-complete description. One possible avenue for exploration is a framework analogous to the $T\bar{T}$-deformation, where the bulk cut-off could be viewed as an IR cut-off (see, e.g., \cite{McGough:2016lol,Guica:2019nzm,Jeong:2019ylz}). While this perspective seems plausible from a gravitational standpoint, its implications for the dual CFT are not yet fully understood. It would be particularly intriguing if this operation could be connected to fuzzball-like states.\footnote{A more conservative approach would be to start with a full AdS black hole geometry and coarse-grain over the interior \cite{Guijosa:2022jdo}. This prescription can also generate an IR cut-off wall, although it is unclear how to engineer the appropriate boundary terms to produce the Gaussian-distributed boundary conditions, or what the CFT interpretation might be.} For a related discussion on the connections between Dirichlet boundary conditions, thermal-like CFT correlators, standard AdS/CFT correlators, and the relationship between black hole quasi-normal modes and normal modes, we refer to~\cite{Krishnan:2023jqn,Burman:2023kko,Banerjee:2024dpl,Banerjee:2024ivh,Burman:2024egy}.}
\item{As a natural extension of our analysis of BTZ black holes, we could apply the brickwall model to describe de Sitter black holes or purely de Sitter spaces. In this context, the stretched horizon could also be placed near the cosmological horizon. Previous studies have explored stretched horizons in de Sitter space \cite{Kim:1998zs, Svesko:2022txo, Banihashemi:2022jys}; however, a complete computation of their normal modes seems to be lacking~\cite{wipfuture}. Investigating these normal modes (along with the analysis of spectral statistics) could provide insights into the distinctions between the von Neumann algebras of black holes \cite{Witten:2021unn} and those of de Sitter space \cite{Chandrasekaran:2022cip}, as well as into the chaotic properties and the phenomenon of hyperfast scrambling in the latter \cite{Susskind:2021esx}.
}
\item{Our analysis has primarily adopted a phenomenological approach. We have demonstrated that specific Dirichlet wall boundary conditions, characterized by particular values of the variance $\sigma_0$, can yield the distinctive features of RMT. However, we did not delve into the underlying conceptual origins of this phenomenon. For a discussion of the relationship between $\sigma_0$ and the level spacing distribution parameter $\beta$ in \eqref{DIS}, please refer to Appendix \ref{appenB}. A deeper exploration of the boundary conditions could provide valuable insights into the relationship between microstate geometries that may emerge in top-down models and the nature of the horizon.}
\item{It would be interesting to investigate whether an effective Hamiltonian and its Hilbert space structure can be defined from the brickwall model, and to explore potential connections to the Eigenstate Thermalization Hypothesis, which concerns not only the spectral statistics but also the statistical properties of energy eigenstates themselves.}
\item{It would be valuable to revisit the intriguing observations made in this paper by exploring more general configurations and other black hole models. For instance, our analysis could be extended to cases involving `massive' scalar and fermionic fields, which may require a more careful examination. The behavior of the AdS boundary expansion can differ depending on whether the difference in conformal dimensions is an integer or half-integer. This distinction implies that normalizability conditions at the asymptotic boundary of AdS may also need to be classified accordingly (for example, see \cite{Natsuume:2020snz} for the scalar case and \cite{Jeong:2023rck} for the fermionic case).}
\item{Additionally, exploring rotating BTZ black holes presents another avenue for extension. It has been shown \cite{Das:2023xjr} that the rotating geometry allows for the consideration of a grand-canonical ensemble for the probe scalar field, along with the corresponding normal modes and the ramp structure in the SFF. A comparative analysis with fermionic fields in this context would also be of significant interest.}
\item{
Finally, investigating higher-dimensional black holes could yield further insights. In \cite{Das:2024fwg}, the normal modes of a probe scalar field are examined within the framework of a five-dimensional AdS-Schwarzschild black hole using a spherically symmetric metric. It would be interesting to study higher-dimensional hyperbolic black holes, as discussed in \cite{Ahn:2020bks, Ahn:2020baf, Ahn:2019rnq}, in order to analytically solve the equations of motion for various fields, including scalar and vector fields, or to consider approximation methods (such as WKB) in more general black hole spacetimes.}
\end{itemize}
We hope to address some these issues in more detail in the near future.

%
\acknowledgments
HSJ and JFP are supported by the Spanish MINECO ‘Centro de Excelencia Severo Ochoa' program under grant SEV-2012-0249, the ‘Atracci\'on de Talento’ program (Comunidad de Madrid) grant 2020-T1/TIC-20495, the Spanish Research Agency via grants CEX2020-001007-S and PID2021-123017NB-I00, funded by MCIN/AEI/10.13039/501100011033, and ERDF `A way of making Europe.' AK is partially supported by DAE-BRNS 58/14/12/2021-BRNS, CRG/2021/004539 and the Department of Atomic Energy, Govt. of India. AK further
acknowledges the support of Humboldt Research Fellowship for Experienced Researchers by the Alexander von Humboldt Foundation.
HSJ would like to thank the Yukawa Institute for Theoretical Physics (YITP), Kyoto University, for their hospitality during the program \textit{STRING DATA 2024}, and the Asia Pacific Center for Theoretical Physics (APCTP), South Korea, for their support during the program \textit{New Avenues in Quantum Many-body Physics and Holography}, where parts of this work were undertaken.

%
\appendix

\section{Dependence of the location of the brickwall}\label{appenA}

In this section, we investigate the dependence of the location of the brickwall or stretched horizon, denoted by $\langle \lambda_J \rangle$. Our primary focus is on the case where $\sigma_0 = 0.025$, which, as demonstrated in the main text, exhibits GUE statistics when $\langle \lambda_J \rangle = -10^4$.

Notably, previous works, such as \cite{Das:2022evy, Das:2023ulz, Das:2023xjr, Krishnan:2023jqn, Krishnan:2024kzf,Krishnan:2024sle}, have shown that signatures of robust chaos -- characterized by the linear ramp in the SFF -- emerge when the stretched horizon is positioned near the black hole. Here, we present novel complementary results by analyzing Krylov complexity in this context.
\begin{figure}[]
 \centering
     \subfigure[$\omega(J=400)$]
     {\includegraphics[width=5.1cm]{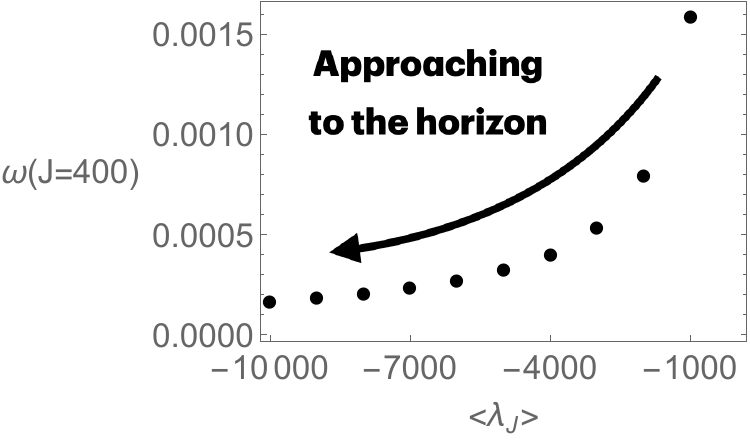} \label{}}
     \subfigure[$C_{\text{peak}}/d$]
     {\includegraphics[width=4.4cm]{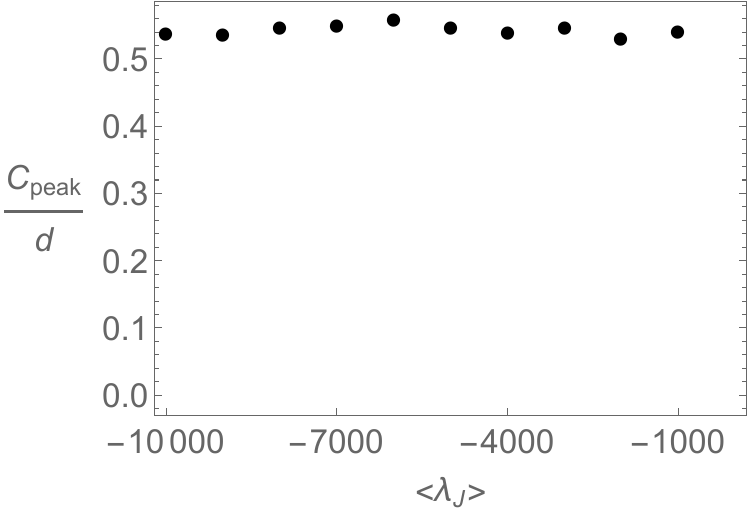} \label{}} 
     \subfigure[$t_{\text{peak}}$]
     {\includegraphics[width=4.8cm]{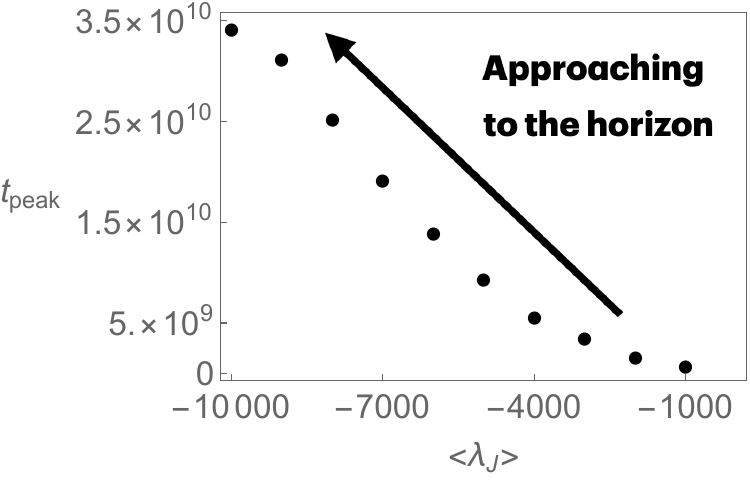} \label{}}
     
     \subfigure[$\omega(J=400)$]
     {\includegraphics[width=5.1cm]{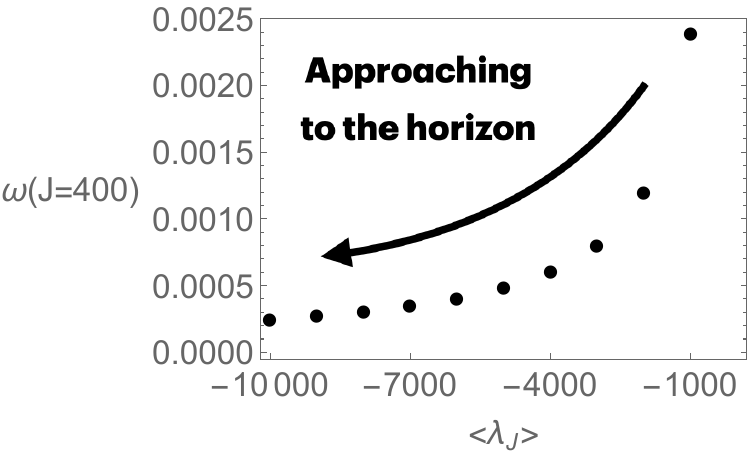} \label{}}
     \subfigure[$C_{\text{peak}}/d$]
     {\includegraphics[width=4.4cm]{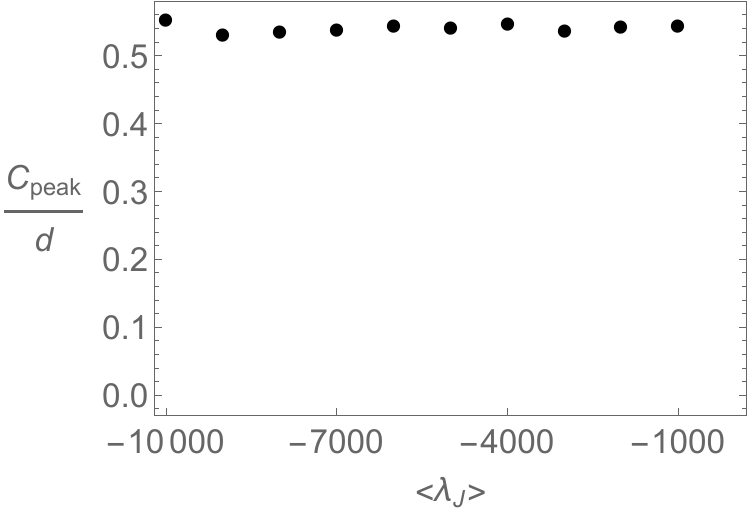} \label{}} 
     \subfigure[$t_{\text{peak}}$]
     {\includegraphics[width=4.8cm]{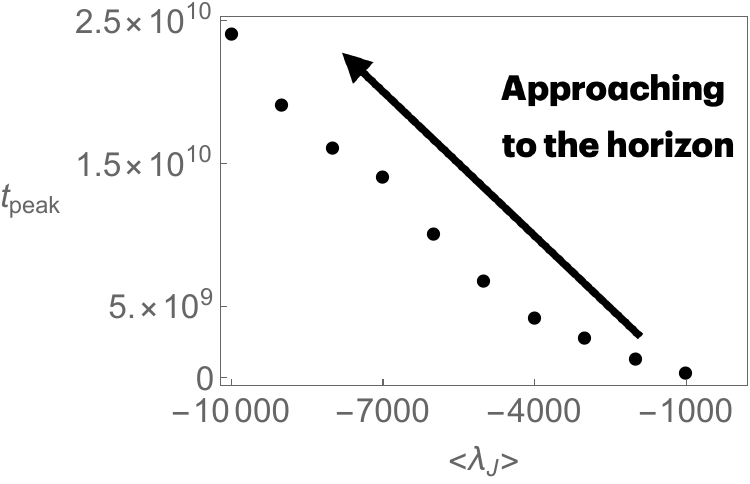} \label{}}
\caption{The dependence of the location of the brickwall: scalar fields (top panel), fermionic fields (bottom panel).}\label{LAMDADEP}
\end{figure}

We begin by examining the normal mode spectrum within the range $\langle \lambda_J \rangle \in [-10^4, -10^3]$, where the stretched horizon approaches the event horizon as $\langle \lambda_J \rangle$ decreases towards $-10^4$. As shown in Figs. \ref{LAMDADEP}(a) and (d), we observe that the normal mode frequencies for a given $J$, such as $\omega(J=400)$, decrease as the stretched horizon nears the event horizon. These findings are consistent with the observations in \cite{Das:2022evy,Das:2023ulz,Das:2023xjr}.
By analyzing the full normal mode spectrum, we confirm that for $\sigma_0 = 0.025$, different values of $\langle \lambda_J \rangle$ also exhibit the GUE level spacing distribution.

Additionally, we study the behavior of Krylov complexity in this regime. We find that a peak in Krylov complexity appears within the range $\langle \lambda_J \rangle \in [-10^4, -10^3]$, as illustrated in Figs. \ref{LAMDADEP}(b) and (e). Moreover, the time scale at which this peak occurs, $t_{\text{peak}}$, increases as the event horizon is approached, as shown in Figs. \ref{LAMDADEP}(c) and (f). This suggests that the peak becomes more prominent and easier to detect (i.e., finite $t_{\text{peak}}$) as the brickwall approaches the event horizon. These results provide a complementary perspective to the SFF analysis conducted in \cite{Das:2022evy,Das:2023ulz,Das:2023xjr}.

\section{Dependence of the variance}\label{appenB}
Here, we analyze the dependence of the variance, $\sigma_0$, in the context of the level spacing distribution, the SFF, and Krylov complexity.

\paragraph{Level spacing distribution.}
We begin by examining the level spacing distribution. To facilitate a more detailed analysis, we introduce the following distribution function\footnote{Here, $\beta$ does not represent the inverse temperature.}:
\begin{align}\label{DIS}
\begin{split}
p(s) = a(\beta, \bar{\beta}) \, s^{\beta} \, \text{exp}\left[ -b(\beta,\bar{\beta}) \, s^{\bar{\beta}+1} \right] \,,
\end{split}
\end{align}
where the coefficients $a(\beta,\bar{\beta})$ and $b(\beta,\bar{\beta})$ are determined by two conditions: normalization to a total probability of $1$, and the requirement that the mean level spacing in units of $s$ is equal to $1$:
\begin{align}
\begin{split}
\int_{0}^{\infty} p(s) \, \dd s =1 \,, \qquad \int_{0}^{\infty} s \, p(s) \, \dd s =1 \,.
\end{split}
\end{align}
These conditions result in 
\begin{align}
\begin{split}
a(\beta,\bar{\beta}) = (1+\bar{\beta}) \frac{\Gamma\left[\frac{2+\beta}{1+\bar{\beta}}\right]^{1+\beta}}{\Gamma\left[\frac{1+\beta}{1+\bar{\beta}}\right]^{2+\beta}} \,, \qquad 
b(\beta,\bar{\beta}) = \frac{\Gamma\left[\frac{2+\beta}{1+\bar{\beta}}\right]^{1+\bar{\beta}}}{\Gamma\left[\frac{1+\beta}{1+\bar{\beta}}\right]^{1+\bar{\beta}}} \,.
\end{split}
\end{align}

Thus, the distribution \eqref{DIS} becomes a function of $(\beta, \bar{\beta})$, allowing it to combine between the Wigner-Dyson with Brody distributions~\cite{Brody1973,Prosen:aa}. This enables the distribution to smoothly transition between various statistical ensembles, including GSE, GUE, GOE, and Poisson statistics. The interpolation mechanism can be described as follows:
\begin{enumerate}
\item{Wigner-Dyson distribution: \eqref{DIS} with $\bar{\beta}=1$ that interpolates between GSE ($\beta=4$), GUE ($\beta=2$), and GOE ($\beta=1$).}
\item{Brody distribution: \eqref{DIS} with $\beta=\bar{\beta}$ that interpolates GOE ($\beta=1$) to Poisson ($\beta=0$).}
\end{enumerate}

\begin{figure}[]
 \centering
     {\includegraphics[width=6.5cm]{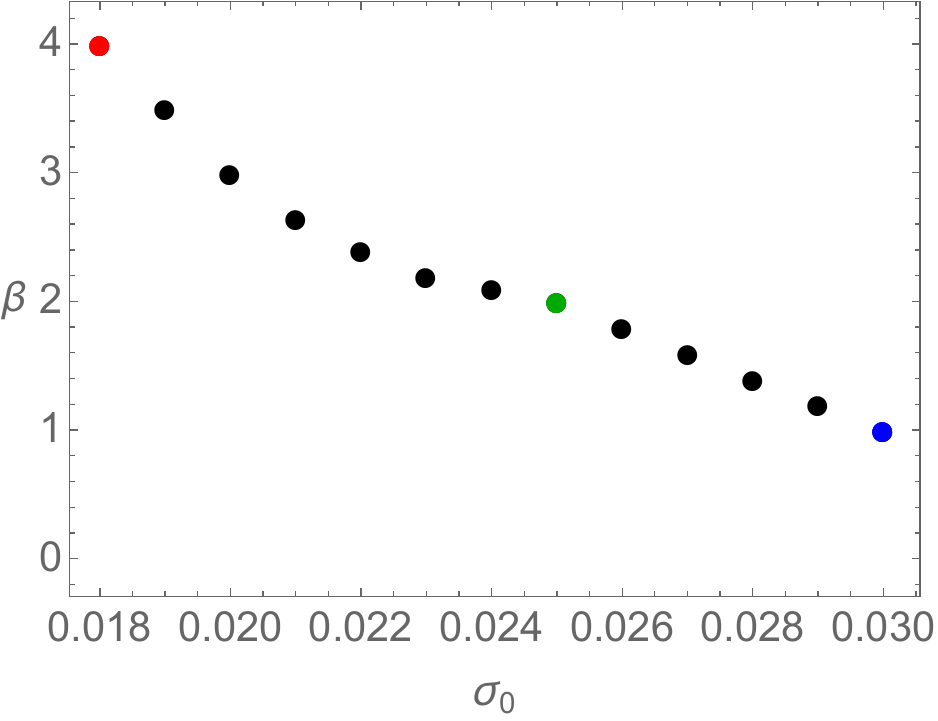} \label{}}
\quad
     {\includegraphics[width=6.8cm]{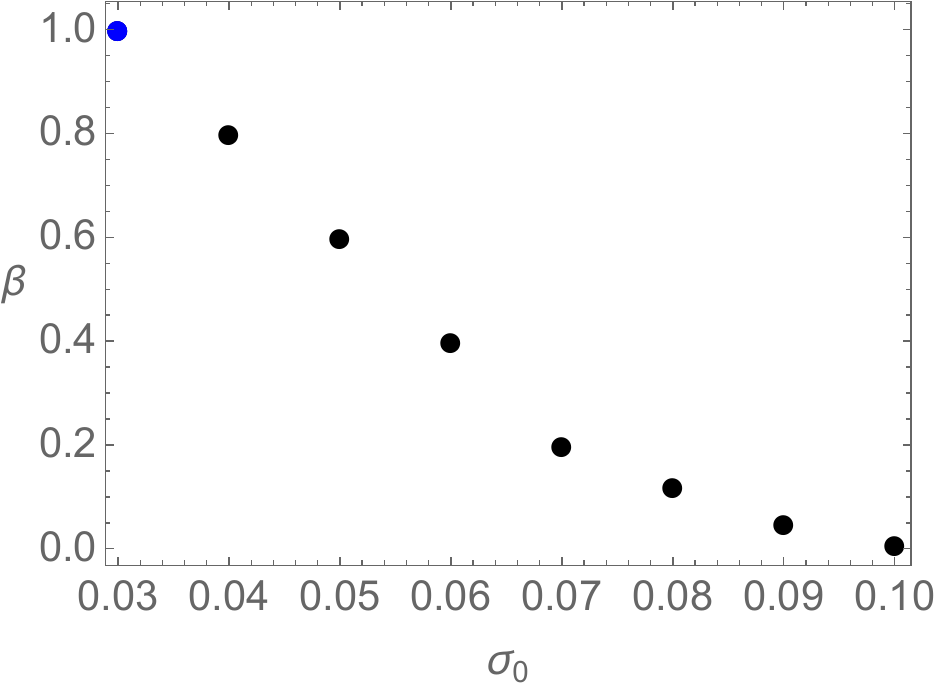} \label{}}
\caption{The level spacing distribution parameter $\beta$ and $\sigma_0$: the Wigner-Dyson distribution (left panel), and the Brody distribution (right panel).}\label{SKET}
\end{figure}

We examine the relationship between $\beta$ and the variance $\sigma_0$, as depicted in Fig. \ref{SKET}. 
We observe that when $\sigma_0 \in [0.018, 0.03]$, the distribution aligns with the Wigner-Dyson distribution: see Fig. \ref{SKET}(a). For values of $\sigma_0 \geq 0.03$, the distribution transitions to the Brody distribution, which interpolates between GOE and Poisson statistics. Specifically, when $0.1 \leq \sigma_0 \leq 2$, the distribution follows Poisson statistics: see Fig. \ref{SKET}(b).

\begin{figure}[]
 \centering
     \subfigure[$\sigma_0 = 0.03$]
     {\includegraphics[width=4.83cm]{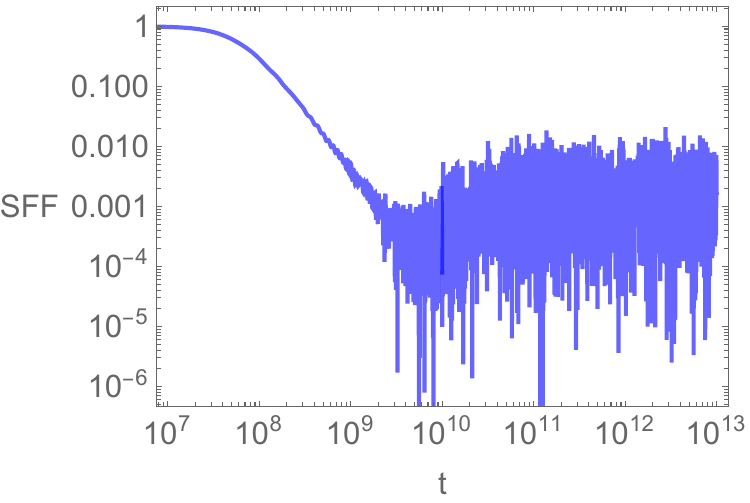} \label{}}
     \subfigure[$\sigma_0 = 0.06$]
     {\includegraphics[width=4.83cm]{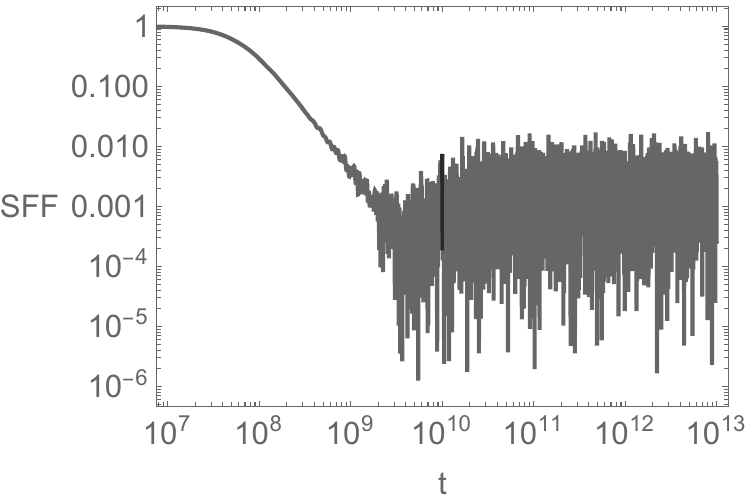} \label{}}
     \subfigure[$\sigma_0 = 0.1$]
     {\includegraphics[width=4.83cm]{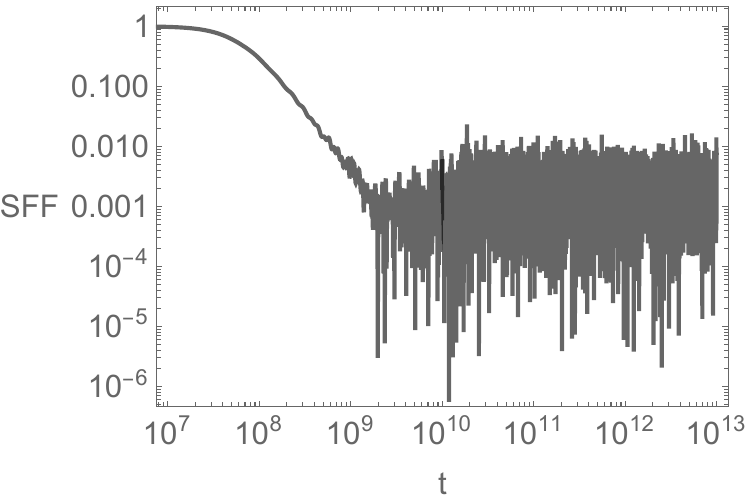} \label{}}
     
     \subfigure[$\sigma_0 = 0.03$]
     {\includegraphics[width=4.83cm]{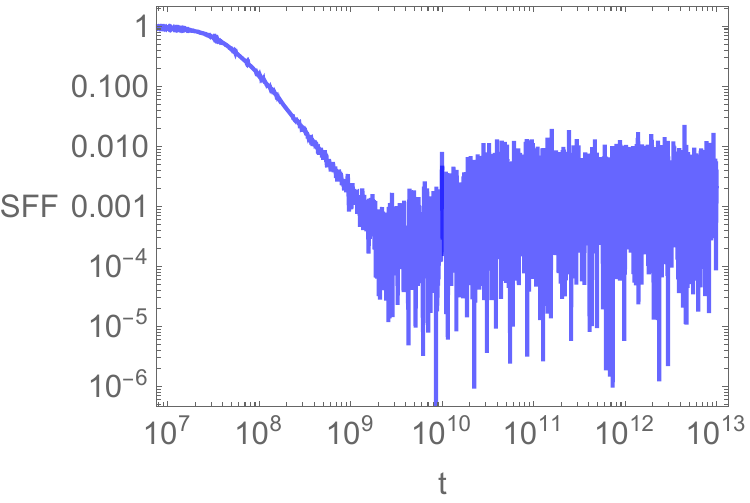} \label{}}
     \subfigure[$\sigma_0 = 0.06$]
     {\includegraphics[width=4.83cm]{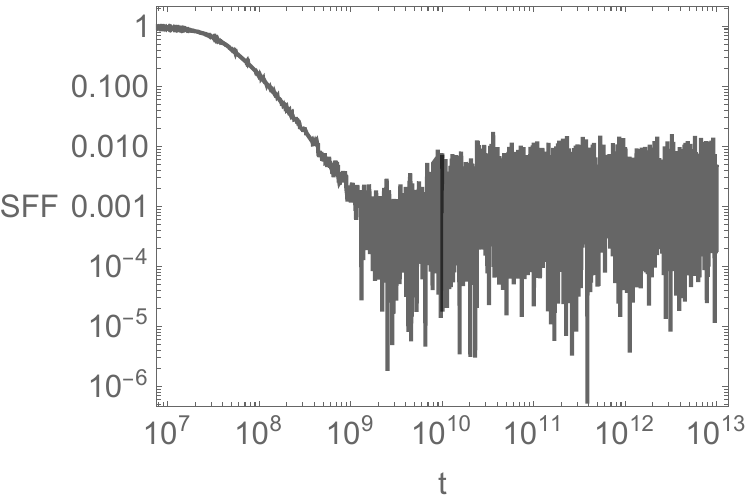} \label{}}
     \subfigure[$\sigma_0 = 0.1$]
     {\includegraphics[width=4.83cm]{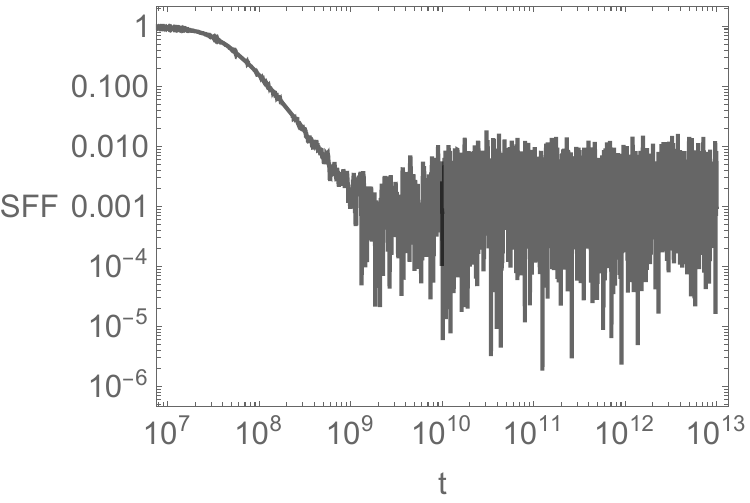} \label{}}     
\caption{SFF of scalar field (top panel), fermionic field (bottom panel).}\label{BRODYSFF}
\end{figure}

\paragraph{Spectral form factor.}
In the case of the SFF, we observe that the characteristic linear ramp in the SFF vanishes as $\sigma_0$ approaches $0.1$, a point at which the Poisson distribution emerges. The transition from GSE to GOE can be seen in Fig. \ref{SFFFIG1}, while the transition from GOE to Poisson is illustrated in Fig. \ref{BRODYSFF}.

\begin{figure}[]
 \centering
     {\includegraphics[width=6.5cm]{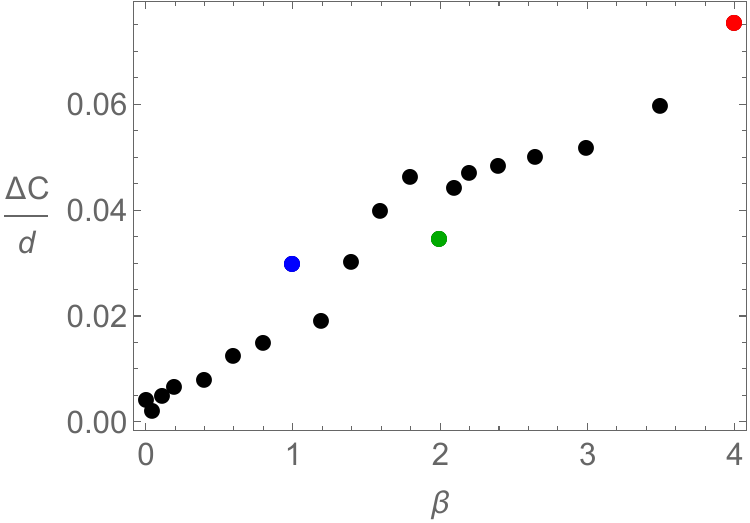} \label{}}
\quad
     {\includegraphics[width=6.5cm]{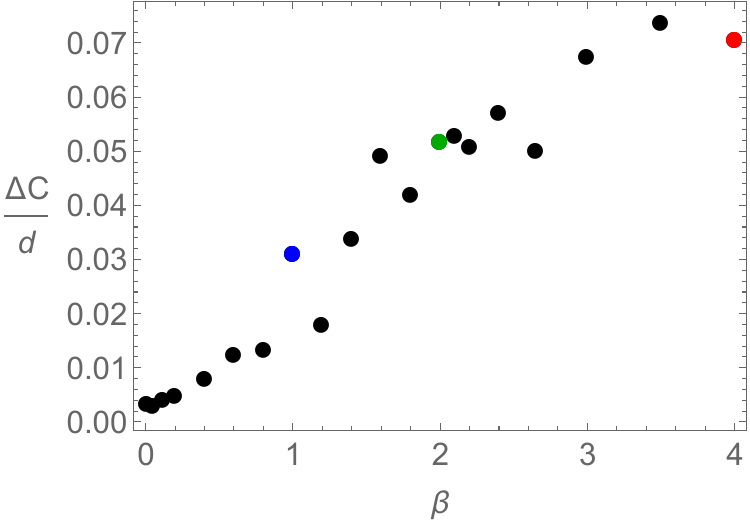} \label{}}
\caption{The peak of Krylov complexity, defined as $\Delta C := C_{\text{peak}} - C(t = \infty)$, as a function of $\beta$, using the expression from \eqref{LATECOMAPP}. The left panel corresponds to the scalar field, while the right panel represents the fermionic field.}\label{KRYFIGSIGMABETA}
\end{figure}

\paragraph{Krylov complexity.}
Finally, we present a plot of the peak of Krylov complexity as a function of $\beta$, or equivalently $\sigma_0$ through in Fig. \ref{SKET}. Our findings indicate that as the distribution approaches the Poisson limit (i.e., $\beta \rightarrow 0$ or $\sigma_0 \rightarrow 0.1$), the characteristic peak of Krylov complexity gradually diminishes. This behavior is illustrated in Fig. \ref{KRYFIGSIGMABETA}.
%

%


\begin{thebibliography}{100}

\bibitem{MichaelBerry_1989}
M.~Berry, \emph{Quantum chaology, not quantum chaos},
  \href{http://dx.doi.org/10.1088/0031-8949/40/3/013}{\emph{Physica Scripta}
  {\bf 40} (sep, 1989) 335}.

\bibitem{D_Alessio_2016}
L.~D'Alessio, Y.~Kafri, A.~Polkovnikov and M.~Rigol, \emph{From quantum chaos
  and eigenstate thermalization to statistical mechanics and thermodynamics},
  \href{http://dx.doi.org/10.1080/00018732.2016.1198134}{\emph{Advances in
  Physics} {\bf 65} (May, 2016) 239--362}.

\bibitem{1977RSPSA.356..375B}
M.~V. {Berry} and M.~{Tabor}, \emph{{Level Clustering in the Regular
  Spectrum}}, \href{http://dx.doi.org/10.1098/rspa.1977.0140}{\emph{Proceedings
  of the Royal Society of London Series A} {\bf 356} (Sept., 1977) 375--394}.

\bibitem{PhysRevLett.52.1}
O.~Bohigas, M.~J. Giannoni and C.~Schmit, \emph{Characterization of chaotic
  quantum spectra and universality of level fluctuation laws},
  \href{http://dx.doi.org/10.1103/PhysRevLett.52.1}{\emph{Phys. Rev. Lett.}
  {\bf 52} (Jan, 1984) 1--4}.

\bibitem{Santos_2010}
L.~F. Santos and M.~Rigol, \emph{Onset of quantum chaos in one-dimensional
  bosonic and fermionic systems and its relation to thermalization},
  \href{http://dx.doi.org/10.1103/physreve.81.036206}{\emph{Physical Review E}
  {\bf 81} (Mar., 2010) }.

\bibitem{DAlessio:2016aa}
L.~D'Alessio, Y.~Kafri, A.~Polkovnikov and M.~Rigol, \emph{From quantum chaos
  and eigenstate thermalization to statistical mechanics and thermodynamics},
  \href{http://dx.doi.org/https://doi.org/10.1080/00018732.2016.1198134}{\emph{Adv.
  Phys.} {\bf 65} (2016) 239}, [\href{http://arxiv.org/abs/1509.06411}{{\tt
  1509.06411}}].

\bibitem{Kobrin:2021aa}
B.~Kobrin, \emph{Many-body chaos in the sachdev-ye-kitaev model},
  \href{http://dx.doi.org/10.1103/PhysRevLett.126.030602}{\emph{Physical Review
  Letters} {\bf 126} (2021) }.

\bibitem{Amore:2024ihm}
P.~Amore, L.~A. Pando~Zayas, J.~F. Pedraza, N.~Quiroz and C.~A.
  Terrero-Escalante, \emph{{Fuzzy Spheres in Stringy Matrix Models: Quantifying
  Chaos in a Mixed Phase Space}},  \href{http://arxiv.org/abs/2407.07259}{{\tt
  2407.07259}}.

\bibitem{Maldacena2016}
J.~Maldacena, S.~H. Shenker and D.~Stanford, \emph{A bound on chaos},
  \href{http://dx.doi.org/10.1007/JHEP08(2016)106}{\emph{Journal of High Energy
  Physics} {\bf 2016} (Aug, 2016) 106}.

\bibitem{Balasubramanian:2022tpr}
V.~Balasubramanian, P.~Caputa, J.~M. Magan and Q.~Wu, \emph{{Quantum chaos and
  the complexity of spread of states}},
  \href{http://dx.doi.org/10.1103/PhysRevD.106.046007}{\emph{Phys. Rev. D} {\bf
  106} (2022) 046007}, [\href{http://arxiv.org/abs/2202.06957}{{\tt
  2202.06957}}].

\bibitem{Parker:2018yvk}
D.~E. Parker, X.~Cao, A.~Avdoshkin, T.~Scaffidi and E.~Altman, \emph{{A
  Universal Operator Growth Hypothesis}},
  \href{http://dx.doi.org/10.1103/PhysRevX.9.041017}{\emph{Phys. Rev. X} {\bf
  9} (2019) 041017}, [\href{http://arxiv.org/abs/1812.08657}{{\tt
  1812.08657}}].

\bibitem{Avdoshkin:2019trj}
A.~Avdoshkin and A.~Dymarsky, \emph{{Euclidean operator growth and quantum
  chaos}},
  \href{http://dx.doi.org/10.1103/PhysRevResearch.2.043234}{\emph{Phys. Rev.
  Res.} {\bf 2} (2020) 043234}, [\href{http://arxiv.org/abs/1911.09672}{{\tt
  1911.09672}}].

\bibitem{Avdoshkin:2022xuw}
A.~Avdoshkin, A.~Dymarsky and M.~Smolkin, \emph{{Krylov complexity in quantum
  field theory, and beyond}},  \href{http://arxiv.org/abs/2212.14429}{{\tt
  2212.14429}}.

\bibitem{Dymarsky:2019elm}
A.~Dymarsky and A.~Gorsky, \emph{{Quantum chaos as delocalization in Krylov
  space}}, \href{http://dx.doi.org/10.1103/PhysRevB.102.085137}{\emph{Phys.
  Rev. B} {\bf 102} (2020) 085137},
  [\href{http://arxiv.org/abs/1912.12227}{{\tt 1912.12227}}].

\bibitem{Dymarsky:2021bjq}
A.~Dymarsky and M.~Smolkin, \emph{{Krylov complexity in conformal field
  theory}}, \href{http://dx.doi.org/10.1103/PhysRevD.104.L081702}{\emph{Phys.
  Rev. D} {\bf 104} (2021) L081702},
  [\href{http://arxiv.org/abs/2104.09514}{{\tt 2104.09514}}].

\bibitem{Chapman:2021jbh}
S.~Chapman and G.~Policastro, \emph{{Quantum computational complexity from
  quantum information to black holes and back}},
  \href{http://dx.doi.org/10.1140/epjc/s10052-022-10037-1}{\emph{Eur. Phys. J.
  C} {\bf 82} (2022) 128}, [\href{http://arxiv.org/abs/2110.14672}{{\tt
  2110.14672}}].

\bibitem{Kundu:2023hbk}
A.~Kundu, V.~Malvimat and R.~Sinha, \emph{{State Dependence of Krylov
  Complexity in $2d$ CFTs}},  \href{http://arxiv.org/abs/2303.03426}{{\tt
  2303.03426}}.

\bibitem{Erdmenger:2023wjg}
J.~Erdmenger, S.-K. Jian and Z.-Y. Xian, \emph{{Universal chaotic dynamics from
  Krylov space}}, \href{http://dx.doi.org/10.1007/JHEP08(2023)176}{\emph{JHEP}
  {\bf 08} (2023) 176}, [\href{http://arxiv.org/abs/2303.12151}{{\tt
  2303.12151}}].

\bibitem{Camargo:2023eev}
H.~A. Camargo, V.~Jahnke, H.-S. Jeong, K.-Y. Kim and M.~Nishida,
  \emph{{Spectral and Krylov Complexity in Billiard Systems}},
  \href{http://arxiv.org/abs/2306.11632}{{\tt 2306.11632}}.

\bibitem{Huh:2023jxt}
K.-B. Huh, H.-S. Jeong and J.~F. Pedraza, \emph{{Spread complexity in
  saddle-dominated scrambling}},
  \href{http://dx.doi.org/10.1007/JHEP05(2024)137}{\emph{JHEP} {\bf 05} (2024)
  137}, [\href{http://arxiv.org/abs/2312.12593}{{\tt 2312.12593}}].

\bibitem{Caputa:2024vrn}
P.~Caputa, H.-S. Jeong, S.~Liu, J.~F. Pedraza and L.-C. Qu, \emph{{Krylov
  complexity of density matrix operators}},
  \href{http://dx.doi.org/10.1007/JHEP05(2024)337}{\emph{JHEP} {\bf 05} (2024)
  337}, [\href{http://arxiv.org/abs/2402.09522}{{\tt 2402.09522}}].

\bibitem{Baggioli:2024wbz}
M.~Baggioli, K.-B. Huh, H.-S. Jeong, K.-Y. Kim and J.~F. Pedraza, \emph{{Krylov
  complexity as an order parameter for quantum chaotic-integrable
  transitions}},  \href{http://arxiv.org/abs/2407.17054}{{\tt 2407.17054}}.

\bibitem{Huh:2024lcm}
K.-B. Huh, H.-S. Jeong, L.~A. Pando~Zayas and J.~F. Pedraza, \emph{{Krylov
  Complexity in Mixed Phase Space}},
  \href{http://arxiv.org/abs/2412.04963}{{\tt 2412.04963}}.

\bibitem{Das:2022evy}
S.~Das, C.~Krishnan, A.~P. Kumar and A.~Kundu, \emph{{Synthetic fuzzballs: a
  linear ramp from black hole normal modes}},
  \href{http://dx.doi.org/10.1007/JHEP01(2023)153}{\emph{JHEP} {\bf 01} (2023)
  153}, [\href{http://arxiv.org/abs/2208.14744}{{\tt 2208.14744}}].

\bibitem{Das:2023ulz}
S.~Das, S.~K. Garg, C.~Krishnan and A.~Kundu, \emph{{Fuzzballs and random
  matrices}}, \href{http://dx.doi.org/10.1007/JHEP10(2023)031}{\emph{JHEP} {\bf
  10} (2023) 031}, [\href{http://arxiv.org/abs/2301.11780}{{\tt 2301.11780}}].

\bibitem{tHooft:1984kcu}
G.~'t~Hooft, \emph{{On the Quantum Structure of a Black Hole}},
  \href{http://dx.doi.org/10.1016/0550-3213(85)90418-3}{\emph{Nucl. Phys. B}
  {\bf 256} (1985) 727--745}.

\bibitem{Kay:2011np}
B.~S. Kay and L.~Ort\'\i{}z, \emph{{Brick Walls and AdS/CFT}},
  \href{http://dx.doi.org/10.1007/s10714-014-1727-x}{\emph{Gen. Rel. Grav.}
  {\bf 46} (2014) 1727}, [\href{http://arxiv.org/abs/1111.6429}{{\tt
  1111.6429}}].

\bibitem{Iizuka:2013kma}
N.~Iizuka and S.~Terashima, \emph{{Brick Walls for Black Holes in AdS/CFT}},
  \href{http://dx.doi.org/10.1016/j.nuclphysb.2015.03.018}{\emph{Nucl. Phys. B}
  {\bf 895} (2015) 1--32}, [\href{http://arxiv.org/abs/1307.5933}{{\tt
  1307.5933}}].

\bibitem{Das:2023xjr}
S.~Das and A.~Kundu, \emph{{Brickwall in rotating BTZ: a dip-ramp-plateau
  story}}, \href{http://dx.doi.org/10.1007/JHEP02(2024)049}{\emph{JHEP} {\bf
  02} (2024) 049}, [\href{http://arxiv.org/abs/2310.06438}{{\tt 2310.06438}}].

\bibitem{Krishnan:2023jqn}
C.~Krishnan and P.~S. Pathak, \emph{{Normal modes of the stretched horizon: a
  bulk mechanism for black hole microstate level spacing}},
  \href{http://dx.doi.org/10.1007/JHEP03(2024)162}{\emph{JHEP} {\bf 03} (2024)
  162}, [\href{http://arxiv.org/abs/2312.14109}{{\tt 2312.14109}}].

\bibitem{Krishnan:2024kzf}
C.~Krishnan and R.~Mondol, \emph{{Young Black Holes Have Smooth Horizons: A
  Swampland Argument}},  \href{http://arxiv.org/abs/2407.11952}{{\tt
  2407.11952}}.

\bibitem{Krishnan:2024sle}
C.~Krishnan and P.~S. Pathak, \emph{{Holomorphic Factorization at the Quantum
  Horizon}},  \href{http://arxiv.org/abs/2410.00732}{{\tt 2410.00732}}.

\bibitem{Denef:2009kn}
F.~Denef, S.~A. Hartnoll and S.~Sachdev, \emph{{Black hole determinants and
  quasinormal modes}},
  \href{http://dx.doi.org/10.1088/0264-9381/27/12/125001}{\emph{Class. Quant.
  Grav.} {\bf 27} (2010) 125001}, [\href{http://arxiv.org/abs/0908.2657}{{\tt
  0908.2657}}].

\bibitem{Mathur:2024mvo}
S.~D. Mathur and M.~Mehta, \emph{{The universal thermodynamic properties of
  Extremely Compact Objects}},  \href{http://arxiv.org/abs/2402.13166}{{\tt
  2402.13166}}.

\bibitem{Brown:2015bva}
A.~R. Brown, D.~A. Roberts, L.~Susskind, B.~Swingle and Y.~Zhao,
  \emph{{Holographic Complexity Equals Bulk Action?}},
  \href{http://dx.doi.org/10.1103/PhysRevLett.116.191301}{\emph{Phys. Rev.
  Lett.} {\bf 116} (2016) 191301}, [\href{http://arxiv.org/abs/1509.07876}{{\tt
  1509.07876}}].

\bibitem{Stanford:2014jda}
D.~Stanford and L.~Susskind, \emph{{Complexity and Shock Wave Geometries}},
  \href{http://dx.doi.org/10.1103/PhysRevD.90.126007}{\emph{Phys. Rev. D} {\bf
  90} (2014) 126007}, [\href{http://arxiv.org/abs/1406.2678}{{\tt 1406.2678}}].

\bibitem{Susskind:2018pmk}
L.~Susskind, \emph{{Three Lectures on Complexity and Black Holes}},
  SpringerBriefs in Physics, Springer, 10, 2018.
\newblock \href{http://arxiv.org/abs/1810.11563}{{\tt 1810.11563}}.
\newblock \href{http://dx.doi.org/10.1007/978-3-030-45109-7}{DOI}.

\bibitem{Rabinovici:2023yex}
E.~Rabinovici, A.~S\'anchez-Garrido, R.~Shir and J.~Sonner, \emph{{A bulk
  manifestation of Krylov complexity}},
  \href{http://dx.doi.org/10.1007/JHEP08(2023)213}{\emph{JHEP} {\bf 08} (2023)
  213}, [\href{http://arxiv.org/abs/2305.04355}{{\tt 2305.04355}}].

\bibitem{Czech:2017ryf}
B.~Czech, \emph{{Einstein Equations from Varying Complexity}},
  \href{http://dx.doi.org/10.1103/PhysRevLett.120.031601}{\emph{Phys. Rev.
  Lett.} {\bf 120} (2018) 031601}, [\href{http://arxiv.org/abs/1706.00965}{{\tt
  1706.00965}}].

\bibitem{Caputa:2018kdj}
P.~Caputa and J.~M. Magan, \emph{{Quantum Computation as Gravity}},
  \href{http://dx.doi.org/10.1103/PhysRevLett.122.231302}{\emph{Phys. Rev.
  Lett.} {\bf 122} (2019) 231302}, [\href{http://arxiv.org/abs/1807.04422}{{\tt
  1807.04422}}].

\bibitem{Pedraza:2021mkh}
J.~F. Pedraza, A.~Russo, A.~Svesko and Z.~Weller-Davies, \emph{{Lorentzian
  Threads as Gatelines and Holographic Complexity}},
  \href{http://dx.doi.org/10.1103/PhysRevLett.127.271602}{\emph{Phys. Rev.
  Lett.} {\bf 127} (2021) 271602}, [\href{http://arxiv.org/abs/2105.12735}{{\tt
  2105.12735}}].

\bibitem{Pedraza:2021fgp}
J.~F. Pedraza, A.~Russo, A.~Svesko and Z.~Weller-Davies, \emph{{Sewing
  spacetime with Lorentzian threads: complexity and the emergence of time in
  quantum gravity}},
  \href{http://dx.doi.org/10.1007/JHEP02(2022)093}{\emph{JHEP} {\bf 02} (2022)
  093}, [\href{http://arxiv.org/abs/2106.12585}{{\tt 2106.12585}}].

\bibitem{Pedraza:2022dqi}
J.~F. Pedraza, A.~Russo, A.~Svesko and Z.~Weller-Davies, \emph{{Computing
  spacetime}}, \href{http://dx.doi.org/10.1142/S021827182242010X}{\emph{Int. J.
  Mod. Phys. D} {\bf 31} (2022) 2242010},
  [\href{http://arxiv.org/abs/2205.05705}{{\tt 2205.05705}}].

\bibitem{Carrasco:2023fcj}
R.~Carrasco, J.~F. Pedraza, A.~Svesko and Z.~Weller-Davies, \emph{{Gravitation
  from optimized computation: Einstein and beyond}},
  \href{http://dx.doi.org/10.1007/JHEP09(2023)167}{\emph{JHEP} {\bf 09} (2023)
  167}, [\href{http://arxiv.org/abs/2306.08503}{{\tt 2306.08503}}].

\bibitem{Lee:2008xf}
S.-S. Lee, \emph{{A Non-Fermi Liquid from a Charged Black Hole: A Critical
  Fermi Ball}},
  \href{http://dx.doi.org/10.1103/PhysRevD.79.086006}{\emph{Phys.Rev.} {\bf
  D79} (2009) 086006}, [\href{http://arxiv.org/abs/0809.3402}{{\tt
  0809.3402}}].

\bibitem{Liu:2009dm}
H.~Liu, J.~McGreevy and D.~Vegh, \emph{{Non-Fermi liquids from holography}},
  \href{http://dx.doi.org/10.1103/PhysRevD.83.065029}{\emph{Phys. Rev.} {\bf
  D83} (2011) 065029}, [\href{http://arxiv.org/abs/0903.2477}{{\tt
  0903.2477}}].

\bibitem{Cubrovic:2009ye}
M.~Cubrovic, J.~Zaanen and K.~Schalm, \emph{{String Theory, Quantum Phase
  Transitions and the Emergent Fermi-Liquid}},
  \href{http://dx.doi.org/10.1126/science.1174962}{\emph{Science} {\bf 325}
  (2009) 439--444}, [\href{http://arxiv.org/abs/0904.1993}{{\tt 0904.1993}}].

\bibitem{Faulkner:2009wj}
T.~Faulkner, H.~Liu, J.~McGreevy and D.~Vegh, \emph{{Emergent quantum
  criticality, Fermi surfaces, and AdS(2)}},
  \href{http://dx.doi.org/10.1103/PhysRevD.83.125002}{\emph{Phys.Rev.} {\bf
  D83} (2011) 125002}, [\href{http://arxiv.org/abs/0907.2694}{{\tt
  0907.2694}}].

\bibitem{Iqbal:2009fd}
N.~Iqbal and H.~Liu, \emph{{Real-time response in AdS/CFT with application to
  spinors}}, \href{http://dx.doi.org/10.1002/prop.200900057}{\emph{Fortsch.
  Phys.} {\bf 57} (2009) 367--384}, [\href{http://arxiv.org/abs/0903.2596}{{\tt
  0903.2596}}].

\bibitem{Dyson:1962es}
F.~J. Dyson, \emph{{Statistical theory of the energy levels of complex systems.
  I}}, \href{http://dx.doi.org/10.1063/1.1703773}{\emph{J. Math. Phys.} {\bf 3}
  (1962) 140--156}.

\bibitem{Bohigas:1983er}
O.~Bohigas, M.~J. Giannoni and C.~Schmit, \emph{{Characterization of chaotic
  quantum spectra and universality of level fluctuation laws}},
  \href{http://dx.doi.org/10.1103/PhysRevLett.52.1}{\emph{Phys. Rev. Lett.}
  {\bf 52} (1984) 1--4}.

\bibitem{Guhr:1997ve}
T.~Guhr, A.~Muller-Groeling and H.~A. Weidenmuller, \emph{{Random matrix
  theories in quantum physics: Common concepts}},
  \href{http://dx.doi.org/10.1016/S0370-1573(97)00088-4}{\emph{Phys. Rept.}
  {\bf 299} (1998) 189--425},
  [\href{http://arxiv.org/abs/cond-mat/9707301}{{\tt cond-mat/9707301}}].

\bibitem{MEHTA1960395}
M.~Mehta, \emph{On the statistical properties of the level-spacings in nuclear
  spectra},
  \href{http://dx.doi.org/https://doi.org/10.1016/0029-5582(60)90413-2}{\emph{Nuclear
  Physics} {\bf 18} (1960) 395--419}.

\bibitem{Dyson:1962oir}
F.~J. Dyson, \emph{{Statistical Theory of the Energy Levels of Complex Systems.
  III}}, \href{http://dx.doi.org/10.1063/1.1703775}{\emph{J. Math. Phys.} {\bf
  3} (1962) 166}.

\bibitem{Wigner_1951}
E.~P. Wigner, \emph{On the statistical distribution of the widths and spacings
  of nuclear resonance levels},
  \href{http://dx.doi.org/10.1017/s0305004100027237}{\emph{Mathematical
  Proceedings of the Cambridge Philosophical Society} {\bf 47} (Oct., 1951)
  790--798}.

\bibitem{Dyson_1962}
F.~J. Dyson, \emph{Statistical theory of the energy levels of complex systems.
  i}, \href{http://dx.doi.org/10.1063/1.1703773}{\emph{Journal of Mathematical
  Physics} {\bf 3} (Jan., 1962) 140--156}.

\bibitem{Brezin:1997aa}
E.~Br{\'e}zin, \emph{Spectral form factor in a random matrix theory},
  \href{http://dx.doi.org/10.1103/PhysRevE.55.4067}{\emph{Physical Review E}
  {\bf 55} (1997) 4067--4083}.

\bibitem{Cotler:2016fpe}
J.~S. Cotler, G.~Gur-Ari, M.~Hanada, J.~Polchinski, P.~Saad, S.~H. Shenker
  et~al., \emph{{Black Holes and Random Matrices}},
  \href{http://dx.doi.org/10.1007/JHEP05(2017)118}{\emph{JHEP} {\bf 05} (2017)
  118}, [\href{http://arxiv.org/abs/1611.04650}{{\tt 1611.04650}}].

\bibitem{Rabinovici:2020ryf}
E.~Rabinovici, A.~S\'anchez-Garrido, R.~Shir and J.~Sonner, \emph{{Operator
  complexity: a journey to the edge of Krylov space}},
  \href{http://dx.doi.org/10.1007/JHEP06(2021)062}{\emph{JHEP} {\bf 06} (2021)
  062}, [\href{http://arxiv.org/abs/2009.01862}{{\tt 2009.01862}}].

\bibitem{Rabinovici:2022beu}
E.~Rabinovici, A.~S\'anchez-Garrido, R.~Shir and J.~Sonner, \emph{{Krylov
  complexity from integrability to chaos}},
  \href{http://dx.doi.org/10.1007/JHEP07(2022)151}{\emph{JHEP} {\bf 07} (2022)
  151}, [\href{http://arxiv.org/abs/2207.07701}{{\tt 2207.07701}}].

\bibitem{Caputa:2022eye}
P.~Caputa and S.~Liu, \emph{{Quantum complexity and topological phases of
  matter}}, \href{http://dx.doi.org/10.1103/PhysRevB.106.195125}{\emph{Phys.
  Rev. B} {\bf 106} (2022) 195125},
  [\href{http://arxiv.org/abs/2205.05688}{{\tt 2205.05688}}].

\bibitem{Afrasiar:2022efk}
M.~Afrasiar, J.~Kumar~Basak, B.~Dey, K.~Pal and K.~Pal, \emph{{Time evolution
  of spread complexity in quenched Lipkin-Meshkov-Glick model}},
  \href{http://arxiv.org/abs/2208.10520}{{\tt 2208.10520}}.

\bibitem{Caputa:2022yju}
P.~Caputa, N.~Gupta, S.~S. Haque, S.~Liu, J.~Murugan and H.~J.~R. Van~Zyl,
  \emph{{Spread complexity and topological transitions in the Kitaev chain}},
  \href{http://dx.doi.org/10.1007/JHEP01(2023)120}{\emph{JHEP} {\bf 01} (2023)
  120}, [\href{http://arxiv.org/abs/2208.06311}{{\tt 2208.06311}}].

\bibitem{Pal:2023yik}
K.~Pal, K.~Pal, A.~Gill and T.~Sarkar, \emph{{Time evolution of spread
  complexity and statistics of work done in quantum quenches}},
  \href{http://arxiv.org/abs/2304.09636}{{\tt 2304.09636}}.

\bibitem{Malvimat:2024vhr}
V.~Malvimat, S.~Porey and B.~Roy, \emph{{Krylov Complexity in $2d$ CFTs with
  SL$(2,\mathbb{R})$ deformed Hamiltonians}},
  \href{http://arxiv.org/abs/2402.15835}{{\tt 2402.15835}}.

\bibitem{Bhattacharjee:2022vlt}
B.~Bhattacharjee, X.~Cao, P.~Nandy and T.~Pathak, \emph{{Krylov complexity in
  saddle-dominated scrambling}},
  \href{http://dx.doi.org/10.1007/JHEP05(2022)174}{\emph{JHEP} {\bf 05} (2022)
  174}, [\href{http://arxiv.org/abs/2203.03534}{{\tt 2203.03534}}].

\bibitem{Bhattacharya:2022gbz}
A.~Bhattacharya, P.~Nandy, P.~P. Nath and H.~Sahu, \emph{{Operator growth and
  Krylov construction in dissipative open quantum systems}},
  \href{http://dx.doi.org/10.1007/JHEP12(2022)081}{\emph{JHEP} {\bf 12} (2022)
  081}, [\href{http://arxiv.org/abs/2207.05347}{{\tt 2207.05347}}].

\bibitem{Bhattacharjee:2022lzy}
B.~Bhattacharjee, X.~Cao, P.~Nandy and T.~Pathak, \emph{{Operator growth in
  open quantum systems: lessons from the dissipative SYK}},
  \href{http://dx.doi.org/10.1007/JHEP03(2023)054}{\emph{JHEP} {\bf 03} (2023)
  054}, [\href{http://arxiv.org/abs/2212.06180}{{\tt 2212.06180}}].

\bibitem{Mohan:2023btr}
V.~Mohan, \emph{{Krylov complexity of open quantum systems: from hard spheres
  to black holes}},
  \href{http://dx.doi.org/10.1007/JHEP11(2023)222}{\emph{JHEP} {\bf 11} (2023)
  222}, [\href{http://arxiv.org/abs/2308.10945}{{\tt 2308.10945}}].

\bibitem{Bhattacharya:2023zqt}
A.~Bhattacharya, P.~Nandy, P.~P. Nath and H.~Sahu, \emph{{On Krylov complexity
  in open systems: an approach via bi-Lanczos algorithm}},
  \href{http://arxiv.org/abs/2303.04175}{{\tt 2303.04175}}.

\bibitem{Bhattacharjee:2023uwx}
B.~Bhattacharjee, P.~Nandy and T.~Pathak, \emph{{Operator dynamics in
  Lindbladian SYK: a Krylov complexity perspective}},
  \href{http://dx.doi.org/10.1007/JHEP01(2024)094}{\emph{JHEP} {\bf 01} (2024)
  094}, [\href{http://arxiv.org/abs/2311.00753}{{\tt 2311.00753}}].

\bibitem{Carolan:2024wov}
E.~Carolan, A.~Kiely, S.~Campbell and S.~Deffner, \emph{{Operator growth and
  spread complexity in open quantum systems}},
  \href{http://dx.doi.org/10.1209/0295-5075/ad5b17}{\emph{EPL} {\bf 147} (2024)
  38002}, [\href{http://arxiv.org/abs/2404.03529}{{\tt 2404.03529}}].

\bibitem{Barbon:2019wsy}
J.~L.~F. Barb\'on, E.~Rabinovici, R.~Shir and R.~Sinha, \emph{{On The Evolution
  Of Operator Complexity Beyond Scrambling}},
  \href{http://dx.doi.org/10.1007/JHEP10(2019)264}{\emph{JHEP} {\bf 10} (2019)
  264}, [\href{http://arxiv.org/abs/1907.05393}{{\tt 1907.05393}}].

\bibitem{Yates:2021asz}
D.~J. Yates and A.~Mitra, \emph{{Strong and almost strong modes of Floquet spin
  chains in Krylov subspaces}},
  \href{http://dx.doi.org/10.1103/PhysRevB.104.195121}{\emph{Phys. Rev. B} {\bf
  104} (2021) 195121}, [\href{http://arxiv.org/abs/2105.13246}{{\tt
  2105.13246}}].

\bibitem{Caputa:2021ori}
P.~Caputa and S.~Datta, \emph{{Operator growth in 2d CFT}},
  \href{http://dx.doi.org/10.1007/JHEP12(2021)188}{\emph{JHEP} {\bf 12} (2021)
  188}, [\href{http://arxiv.org/abs/2110.10519}{{\tt 2110.10519}}].

\bibitem{Patramanis:2021lkx}
D.~Patramanis, \emph{{Probing the entanglement of operator growth}},
  \href{http://dx.doi.org/10.1093/ptep/ptac081}{\emph{PTEP} {\bf 2022} (2022)
  063A01}, [\href{http://arxiv.org/abs/2111.03424}{{\tt 2111.03424}}].

\bibitem{Trigueros:2021rwj}
F.~B. Trigueros and C.-J. Lin, \emph{{Krylov complexity of many-body
  localization: Operator localization in Krylov basis}},
  \href{http://arxiv.org/abs/2112.04722}{{\tt 2112.04722}}.

\bibitem{Rabinovici:2021qqt}
E.~Rabinovici, A.~S\'anchez-Garrido, R.~Shir and J.~Sonner, \emph{{Krylov
  localization and suppression of complexity}},
  \href{http://dx.doi.org/10.1007/JHEP03(2022)211}{\emph{JHEP} {\bf 03} (2022)
  211}, [\href{http://arxiv.org/abs/2112.12128}{{\tt 2112.12128}}].

\bibitem{Bhattacharya:2023xjx}
A.~Bhattacharya, P.~P. Nath and H.~Sahu, \emph{{Krylov complexity for nonlocal
  spin chains}},
  \href{http://dx.doi.org/10.1103/PhysRevD.109.066010}{\emph{Phys. Rev. D} {\bf
  109} (2024) 066010}, [\href{http://arxiv.org/abs/2312.11677}{{\tt
  2312.11677}}].

\bibitem{Bhattacharjee:2022qjw}
B.~Bhattacharjee, S.~Sur and P.~Nandy, \emph{{Probing quantum scars and weak
  ergodicity breaking through quantum complexity}},
  \href{http://dx.doi.org/10.1103/PhysRevB.106.205150}{\emph{Phys. Rev. B} {\bf
  106} (2022) 205150}, [\href{http://arxiv.org/abs/2208.05503}{{\tt
  2208.05503}}].

\bibitem{Chattopadhyay:2023fob}
A.~Chattopadhyay, A.~Mitra and H.~J.~R. van Zyl, \emph{{Spread complexity as
  classical dilaton solutions}},
  \href{http://dx.doi.org/10.1103/PhysRevD.108.025013}{\emph{Phys. Rev. D} {\bf
  108} (2023) 025013}, [\href{http://arxiv.org/abs/2302.10489}{{\tt
  2302.10489}}].

\bibitem{Bhattacharjee:2023dik}
B.~Bhattacharjee, \emph{{A Lanczos approach to the Adiabatic Gauge Potential}},
   \href{http://arxiv.org/abs/2302.07228}{{\tt 2302.07228}}.

\bibitem{Bhattacharjee:2022ave}
B.~Bhattacharjee, P.~Nandy and T.~Pathak, \emph{{Krylov complexity in large q
  and double-scaled SYK model}},
  \href{http://dx.doi.org/10.1007/JHEP08(2023)099}{\emph{JHEP} {\bf 08} (2023)
  099}, [\href{http://arxiv.org/abs/2210.02474}{{\tt 2210.02474}}].

\bibitem{Takahashi:2023nkt}
K.~Takahashi and A.~del Campo, \emph{{Shortcuts to Adiabaticity in Krylov
  Space}}, \href{http://dx.doi.org/10.1103/PhysRevX.14.011032}{\emph{Phys. Rev.
  X} {\bf 14} (2024) 011032}, [\href{http://arxiv.org/abs/2302.05460}{{\tt
  2302.05460}}].

\bibitem{Camargo:2022rnt}
H.~A. Camargo, V.~Jahnke, K.-Y. Kim and M.~Nishida, \emph{{Krylov complexity in
  free and interacting scalar field theories with bounded power spectrum}},
  \href{http://dx.doi.org/10.1007/JHEP05(2023)226}{\emph{JHEP} {\bf 05} (2023)
  226}, [\href{http://arxiv.org/abs/2212.14702}{{\tt 2212.14702}}].

\bibitem{Hashimoto:2023swv}
K.~Hashimoto, K.~Murata, N.~Tanahashi and R.~Watanabe, \emph{{Krylov complexity
  and chaos in quantum mechanics}},
  \href{http://dx.doi.org/10.1007/JHEP11(2023)040}{\emph{JHEP} {\bf 11} (2023)
  040}, [\href{http://arxiv.org/abs/2305.16669}{{\tt 2305.16669}}].

\bibitem{Iizuka:2023pov}
N.~Iizuka and M.~Nishida, \emph{{Krylov complexity in the IP matrix model}},
  \href{http://dx.doi.org/10.1007/JHEP11(2023)065}{\emph{JHEP} {\bf 11} (2023)
  065}, [\href{http://arxiv.org/abs/2306.04805}{{\tt 2306.04805}}].

\bibitem{Caputa:2023vyr}
P.~Caputa, J.~M. Magan, D.~Patramanis and E.~Tonni, \emph{{Krylov complexity of
  modular Hamiltonian evolution}},
  \href{http://dx.doi.org/10.1103/PhysRevD.109.086004}{\emph{Phys. Rev. D} {\bf
  109} (2024) 086004}, [\href{http://arxiv.org/abs/2306.14732}{{\tt
  2306.14732}}].

\bibitem{Fan:2023ohh}
Z.-Y. Fan, \emph{{Generalised Krylov complexity}},
  \href{http://arxiv.org/abs/2306.16118}{{\tt 2306.16118}}.

\bibitem{Vasli:2023syq}
M.~J. Vasli, K.~Babaei~Velni, M.~R. Mohammadi~Mozaffar, A.~Mollabashi and
  M.~Alishahiha, \emph{{Krylov complexity in Lifshitz-type scalar field
  theories}},
  \href{http://dx.doi.org/10.1140/epjc/s10052-024-12609-9}{\emph{Eur. Phys. J.
  C} {\bf 84} (2024) 235}, [\href{http://arxiv.org/abs/2307.08307}{{\tt
  2307.08307}}].

\bibitem{Gautam:2023bcm}
M.~Gautam, K.~Pal, K.~Pal, A.~Gill, N.~Jaiswal and T.~Sarkar, \emph{{Spread
  complexity evolution in quenched interacting quantum systems}},
  \href{http://arxiv.org/abs/2308.00636}{{\tt 2308.00636}}.

\bibitem{Iizuka:2023fba}
N.~Iizuka and M.~Nishida, \emph{{Krylov complexity in the IP matrix model. Part
  II}}, \href{http://dx.doi.org/10.1007/JHEP11(2023)096}{\emph{JHEP} {\bf 11}
  (2023) 096}, [\href{http://arxiv.org/abs/2308.07567}{{\tt 2308.07567}}].

\bibitem{Anegawa:2024wov}
T.~Anegawa, N.~Iizuka and M.~Nishida, \emph{{Krylov complexity as an order
  parameter for deconfinement phase transitions at large N}},
  \href{http://dx.doi.org/10.1007/JHEP04(2024)119}{\emph{JHEP} {\bf 04} (2024)
  119}, [\href{http://arxiv.org/abs/2401.04383}{{\tt 2401.04383}}].

\bibitem{Chen:2024imd}
L.~Chen, B.~Mu, H.~Wang and P.~Zhang, \emph{{Dissecting Quantum Many-body Chaos
  in the Krylov Space}},  \href{http://arxiv.org/abs/2404.08207}{{\tt
  2404.08207}}.

\bibitem{Caputa:2024xkp}
P.~Caputa and K.~Kutak, \emph{{Krylov complexity and gluon cascades in the high
  energy limit}},
  \href{http://dx.doi.org/10.1103/PhysRevD.110.085011}{\emph{Phys. Rev. D} {\bf
  110} (2024) 085011}, [\href{http://arxiv.org/abs/2404.07657}{{\tt
  2404.07657}}].

\bibitem{Chattopadhyay:2024pdj}
A.~Chattopadhyay, V.~Malvimat and A.~Mitra, \emph{{Krylov complexity of
  deformed conformal field theories}},
  \href{http://dx.doi.org/10.1007/JHEP08(2024)053}{\emph{JHEP} {\bf 08} (2024)
  053}, [\href{http://arxiv.org/abs/2405.03630}{{\tt 2405.03630}}].

\bibitem{Nandy:2023brt}
S.~Nandy, B.~Mukherjee, A.~Bhattacharyya and A.~Banerjee, \emph{{Quantum state
  complexity meets many-body scars}},
  \href{http://dx.doi.org/10.1088/1361-648X/ad1a7b}{\emph{J. Phys. Condens.
  Matter} {\bf 36} (2024) 155601}, [\href{http://arxiv.org/abs/2305.13322}{{\tt
  2305.13322}}].

\bibitem{Aguilar-Gutierrez:2023nyk}
S.~E. Aguilar-Gutierrez and A.~Rolph, \emph{{Krylov complexity is not a measure
  of distance between states or operators}},
  \href{http://dx.doi.org/10.1103/PhysRevD.109.L081701}{\emph{Phys. Rev. D}
  {\bf 109} (2024) L081701}, [\href{http://arxiv.org/abs/2311.04093}{{\tt
  2311.04093}}].

\bibitem{Camargo:2024deu}
H.~A. Camargo, K.-B. Huh, V.~Jahnke, H.-S. Jeong, K.-Y. Kim and M.~Nishida,
  \emph{{Spread and spectral complexity in quantum spin chains: from
  integrability to chaos}},
  \href{http://dx.doi.org/10.1007/JHEP08(2024)241}{\emph{JHEP} {\bf 08} (2024)
  241}, [\href{http://arxiv.org/abs/2405.11254}{{\tt 2405.11254}}].

\bibitem{Aguilar-Gutierrez:2024nau}
S.~E. Aguilar-Gutierrez, \emph{{Towards complexity in de Sitter space from the
  double-scaled Sachdev-Ye-Kitaev model}},
  \href{http://arxiv.org/abs/2403.13186}{{\tt 2403.13186}}.

\bibitem{Bhattacharjee:2024yxj}
B.~Bhattacharjee and P.~Nandy, \emph{{Krylov fractality and complexity in
  generic random matrix ensembles}},
  \href{http://arxiv.org/abs/2407.07399}{{\tt 2407.07399}}.

\bibitem{Balasubramanian:2024ghv}
V.~Balasubramanian, R.~N. Das, J.~Erdmenger and Z.-Y. Xian, \emph{{Chaos and
  integrability in triangular billiards}},
  \href{http://arxiv.org/abs/2407.11114}{{\tt 2407.11114}}.

\bibitem{Craps:2024suj}
B.~Craps, O.~Evnin and G.~Pascuzzi, \emph{{Multiseed Krylov complexity}},
  \href{http://arxiv.org/abs/2409.15666}{{\tt 2409.15666}}.

\bibitem{Alishahiha:2024vbf}
M.~Alishahiha, S.~Banerjee and M.~J. Vasli, \emph{{Krylov Complexity as a Probe
  for Chaos}},  \href{http://arxiv.org/abs/2408.10194}{{\tt 2408.10194}}.

\bibitem{Gill:2024acg}
A.~Gill and T.~Sarkar, \emph{{Speed Limits and Scrambling in Krylov Space}},
  \href{http://arxiv.org/abs/2408.06855}{{\tt 2408.06855}}.

\bibitem{Li:2024ljz}
T.~Li and L.-H. Liu, \emph{{Krylov complexity of thermal state in early
  universe}},  \href{http://arxiv.org/abs/2408.03293}{{\tt 2408.03293}}.

\bibitem{Jha:2024nbl}
R.~G. Jha and R.~Roy, \emph{{Sparsity dependence of Krylov state complexity in
  the SYK model}},  \href{http://arxiv.org/abs/2407.20569}{{\tt 2407.20569}}.

\bibitem{Camargo:2024rrj}
H.~A. Camargo, Y.~Fu, V.~Jahnke, K.-Y. Kim and K.~Pal, \emph{{Higher-Order
  Krylov State Complexity in Random Matrix Quenches}},
  \href{http://arxiv.org/abs/2412.16472}{{\tt 2412.16472}}.

\bibitem{He:2022ryk}
S.~He, P.~H.~C. Lau, Z.-Y. Xian and L.~Zhao, \emph{{Quantum chaos, scrambling
  and operator growth in $ T\overline{T} $ deformed SYK models}},
  \href{http://dx.doi.org/10.1007/JHEP12(2022)070}{\emph{JHEP} {\bf 12} (2022)
  070}, [\href{http://arxiv.org/abs/2209.14936}{{\tt 2209.14936}}].

\bibitem{Nandy:2024mml}
P.~Nandy, T.~Pathak, Z.-Y. Xian and J.~Erdmenger, \emph{{A Krylov space
  approach to Singular Value Decomposition in non-Hermitian systems}},
  \href{http://arxiv.org/abs/2411.09309}{{\tt 2411.09309}}.

\bibitem{Li:2024kfm}
T.~Li and L.-H. Liu, \emph{{Inflationary Krylov complexity}},
  \href{http://dx.doi.org/10.1007/JHEP04(2024)123}{\emph{JHEP} {\bf 04} (2024)
  123}, [\href{http://arxiv.org/abs/2401.09307}{{\tt 2401.09307}}].

\bibitem{Zhai:2024odw}
K.-H. Zhai and L.-H. Liu, \emph{{Krylov Complexity in early universe}},
  \href{http://arxiv.org/abs/2411.18405}{{\tt 2411.18405}}.

\bibitem{Zhai:2024tkz}
K.-H. Zhai, L.-H. Liu and H.-Q. Zhang, \emph{{The generalized CV conjecture of
  Krylov complexity}},  \href{http://arxiv.org/abs/2412.08925}{{\tt
  2412.08925}}.

\bibitem{Xu:2024gfm}
J.~Xu, \emph{{On Chord Dynamics and Complexity Growth in Double-Scaled SYK}},
  \href{http://arxiv.org/abs/2411.04251}{{\tt 2411.04251}}.

\bibitem{Bhattacharya:2024szw}
A.~Bhattacharya and A.~Jana, \emph{{Quantum chaos and complexity from string
  scattering amplitudes}},  \href{http://arxiv.org/abs/2408.11096}{{\tt
  2408.11096}}.

\bibitem{Nandy:2024htc}
P.~Nandy, A.~S. Matsoukas-Roubeas, P.~Mart\'\i{}nez-Azcona, A.~Dymarsky and
  A.~del Campo, \emph{{Quantum Dynamics in Krylov Space: Methods and
  Applications}},  \href{http://arxiv.org/abs/2405.09628}{{\tt 2405.09628}}.

\bibitem{Lanczos:1950zz}
C.~Lanczos, \emph{{An iteration method for the solution of the eigenvalue
  problem of linear differential and integral operators}},
  \href{http://dx.doi.org/10.6028/jres.045.026}{\emph{J. Res. Natl. Bur. Stand.
  B} {\bf 45} (1950) 255--282}.

\bibitem{Banados:1992wn}
M.~Banados, C.~Teitelboim and J.~Zanelli, \emph{{The Black hole in
  three-dimensional space-time}},
  \href{http://dx.doi.org/10.1103/PhysRevLett.69.1849}{\emph{Phys. Rev. Lett.}
  {\bf 69} (1992) 1849--1851}, [\href{http://arxiv.org/abs/hep-th/9204099}{{\tt
  hep-th/9204099}}].

\bibitem{Banados:1992gq}
M.~Banados, M.~Henneaux, C.~Teitelboim and J.~Zanelli, \emph{{Geometry of the
  (2+1) black hole}},
  \href{http://dx.doi.org/10.1103/PhysRevD.48.1506}{\emph{Phys. Rev. D} {\bf
  48} (1993) 1506--1525}, [\href{http://arxiv.org/abs/gr-qc/9302012}{{\tt
  gr-qc/9302012}}].

\bibitem{Das:2023yfj}
S.~Das, S.~K. Garg, C.~Krishnan and A.~Kundu, \emph{{What is the Simplest
  Linear Ramp?}}, \href{http://dx.doi.org/10.1007/JHEP01(2024)172}{\emph{JHEP}
  {\bf 01} (2024) 172}, [\href{http://arxiv.org/abs/2308.11704}{{\tt
  2308.11704}}].

\bibitem{He:2024hkw}
P.-Z. He and H.-Q. Zhang, \emph{{Krylov Complexity in the Schr\"odinger Field
  Theory}},  \href{http://arxiv.org/abs/2411.16302}{{\tt 2411.16302}}.

\bibitem{McGough:2016lol}
L.~McGough, M.~Mezei and H.~Verlinde, \emph{{Moving the CFT into the bulk with
  $ T\overline{T} $}},
  \href{http://dx.doi.org/10.1007/JHEP04(2018)010}{\emph{JHEP} {\bf 04} (2018)
  010}, [\href{http://arxiv.org/abs/1611.03470}{{\tt 1611.03470}}].

\bibitem{Guica:2019nzm}
M.~Guica and R.~Monten, \emph{{$T\bar T$ and the mirage of a bulk cutoff}},
  \href{http://dx.doi.org/10.21468/SciPostPhys.10.2.024}{\emph{SciPost Phys.}
  {\bf 10} (2021) 024}, [\href{http://arxiv.org/abs/1906.11251}{{\tt
  1906.11251}}].

\bibitem{Jeong:2019ylz}
H.-S. Jeong, K.-Y. Kim and M.~Nishida, \emph{{Entanglement and R\'enyi entropy
  of multiple intervals in $T\overline{T}$-deformed CFT and holography}},
  \href{http://dx.doi.org/10.1103/PhysRevD.100.106015}{\emph{Phys. Rev. D} {\bf
  100} (2019) 106015}, [\href{http://arxiv.org/abs/1906.03894}{{\tt
  1906.03894}}].

\bibitem{Guijosa:2022jdo}
A.~Guijosa, Y.~D. Olivas and J.~F. Pedraza, \emph{{Holographic coarse-graining:
  correlators from the entanglement wedge and other reduced geometries}},
  \href{http://dx.doi.org/10.1007/JHEP08(2022)118}{\emph{JHEP} {\bf 08} (2022)
  118}, [\href{http://arxiv.org/abs/2201.01786}{{\tt 2201.01786}}].

\bibitem{Burman:2023kko}
V.~Burman, S.~Das and C.~Krishnan, \emph{{A smooth horizon without a smooth
  horizon}}, \href{http://dx.doi.org/10.1007/JHEP03(2024)014}{\emph{JHEP} {\bf
  2024} (2024) 014}, [\href{http://arxiv.org/abs/2312.14108}{{\tt
  2312.14108}}].

\bibitem{Banerjee:2024dpl}
S.~Banerjee, S.~Das, M.~Dorband and A.~Kundu, \emph{{Brickwall, normal modes,
  and emerging thermality}},
  \href{http://dx.doi.org/10.1103/PhysRevD.109.126020}{\emph{Phys. Rev. D} {\bf
  109} (2024) 126020}, [\href{http://arxiv.org/abs/2401.01417}{{\tt
  2401.01417}}].

\bibitem{Banerjee:2024ivh}
S.~Banerjee, S.~Das, A.~Kundu and M.~Sittinger, \emph{{Blackish Holes}},
  \href{http://arxiv.org/abs/2411.09500}{{\tt 2411.09500}}.

\bibitem{Burman:2024egy}
V.~Burman and C.~Krishnan, \emph{{A Bottom-Up Approach to Black Hole
  Microstates}},  \href{http://arxiv.org/abs/2409.05850}{{\tt 2409.05850}}.

\bibitem{Kim:1998zs}
W.~T. Kim, \emph{{Entropy of (2+1)-dimensional de Sitter space in terms of
  brick wall method}},
  \href{http://dx.doi.org/10.1103/PhysRevD.59.047503}{\emph{Phys. Rev. D} {\bf
  59} (1999) 047503}, [\href{http://arxiv.org/abs/hep-th/9810169}{{\tt
  hep-th/9810169}}].

\bibitem{Svesko:2022txo}
A.~Svesko, E.~Verheijden, E.~P. Verlinde and M.~R. Visser, \emph{{Quasi-local
  energy and microcanonical entropy in two-dimensional nearly de Sitter
  gravity}}, \href{http://dx.doi.org/10.1007/JHEP08(2022)075}{\emph{JHEP} {\bf
  08} (2022) 075}, [\href{http://arxiv.org/abs/2203.00700}{{\tt 2203.00700}}].

\bibitem{Banihashemi:2022jys}
B.~Banihashemi and T.~Jacobson, \emph{{Thermodynamic ensembles with
  cosmological horizons}},
  \href{http://dx.doi.org/10.1007/JHEP07(2022)042}{\emph{JHEP} {\bf 07} (2022)
  042}, [\href{http://arxiv.org/abs/2204.05324}{{\tt 2204.05324}}].

\bibitem{wipfuture}
H.-S. Jeong, J.~F. Pedraza and J.~M. Begines, \emph{work in progress},
  {\emph{to appear} }.

\bibitem{Witten:2021unn}
E.~Witten, \emph{{Gravity and the crossed product}},
  \href{http://dx.doi.org/10.1007/JHEP10(2022)008}{\emph{JHEP} {\bf 10} (2022)
  008}, [\href{http://arxiv.org/abs/2112.12828}{{\tt 2112.12828}}].

\bibitem{Chandrasekaran:2022cip}
V.~Chandrasekaran, R.~Longo, G.~Penington and E.~Witten, \emph{{An algebra of
  observables for de Sitter space}},
  \href{http://dx.doi.org/10.1007/JHEP02(2023)082}{\emph{JHEP} {\bf 02} (2023)
  082}, [\href{http://arxiv.org/abs/2206.10780}{{\tt 2206.10780}}].

\bibitem{Susskind:2021esx}
L.~Susskind, \emph{{Entanglement and Chaos in De Sitter Space Holography: An
  SYK Example}},
  \href{http://dx.doi.org/10.22128/jhap.2021.455.1005}{\emph{JHAP} {\bf 1}
  (2021) 1--22}, [\href{http://arxiv.org/abs/2109.14104}{{\tt 2109.14104}}].

\bibitem{Natsuume:2020snz}
M.~Natsuume and T.~Okamura, \emph{{Pole-skipping and zero temperature}},
  \href{http://dx.doi.org/10.1103/PhysRevD.103.066017}{\emph{Phys. Rev. D} {\bf
  103} (2021) 066017}, [\href{http://arxiv.org/abs/2011.10093}{{\tt
  2011.10093}}].

\bibitem{Jeong:2023rck}
H.-S. Jeong, C.-W. Ji and K.-Y. Kim, \emph{{Pole-skipping in rotating BTZ black
  holes}}, \href{http://dx.doi.org/10.1007/JHEP08(2023)139}{\emph{JHEP} {\bf
  08} (2023) 139}, [\href{http://arxiv.org/abs/2306.14805}{{\tt 2306.14805}}].

\bibitem{Das:2024fwg}
S.~Das, S.~Porey and B.~Roy, \emph{{Brick Wall in AdS-Schwarzschild Black Hole:
  Normal Modes and Emerging Thermality}},
  \href{http://arxiv.org/abs/2409.05519}{{\tt 2409.05519}}.

\bibitem{Ahn:2020bks}
Y.~Ahn, V.~Jahnke, H.-S. Jeong, K.-Y. Kim, K.-S. Lee and M.~Nishida,
  \emph{{Pole-skipping of scalar and vector fields in hyperbolic space:
  conformal blocks and holography}},
  \href{http://dx.doi.org/10.1007/JHEP09(2020)111}{\emph{JHEP} {\bf 09} (2020)
  111}, [\href{http://arxiv.org/abs/2006.00974}{{\tt 2006.00974}}].

\bibitem{Ahn:2020baf}
Y.~Ahn, V.~Jahnke, H.-S. Jeong, K.-Y. Kim, K.-S. Lee and M.~Nishida,
  \emph{{Classifying pole-skipping points}},
  \href{http://dx.doi.org/10.1007/JHEP03(2021)175}{\emph{JHEP} {\bf 03} (2021)
  175}, [\href{http://arxiv.org/abs/2010.16166}{{\tt 2010.16166}}].

\bibitem{Ahn:2019rnq}
Y.~Ahn, V.~Jahnke, H.-S. Jeong and K.-Y. Kim, \emph{{Scrambling in Hyperbolic
  Black Holes: shock waves and pole-skipping}},
  \href{http://dx.doi.org/10.1007/JHEP10(2019)257}{\emph{JHEP} {\bf 10} (2019)
  257}, [\href{http://arxiv.org/abs/1907.08030}{{\tt 1907.08030}}].

\bibitem{Brody1973}
T.~A. Brody, \emph{A statistical measure for the repulsion of energy levels},
  \href{http://dx.doi.org/10.1007/BF02727859}{\emph{Lettere al Nuovo Cimento
  (1971-1985)} {\bf 7} (Jul, 1973) 482--484}.

\bibitem{Prosen:aa}
T.~Prosen, \emph{Berry-robnik level statistics in a smooth billiard system},
  \href{http://arxiv.org/abs/cond-mat/9803341}{{\tt cond-mat/9803341}}.

\end{thebibliography}
\bibliographystyle{JHEP}

\providecommand{\href}[2]{#2}\begingroup\raggedright\endgroup

\end{document}